\documentclass[pra,aps,eqsecnum,showpacs,superscriptaddress]{revtex4-1}
\usepackage{amsmath}
\usepackage{amssymb}
\usepackage{latexsym}
\usepackage{rotating}
\usepackage{enumerate}
\usepackage{changepage}
\usepackage{float}
\usepackage{graphicx}
\newtheorem{theorem}{Theorem}

\newcommand{\ket} [1] {\vert #1 \rangle}
\newcommand{\bra} [1] {\langle #1 \vert}

\newcommand{\proj}[1]{\ket{#1}\bra{#1}}

\newcommand{\hr}[1]{\hat{\rho}_{#1}}

\begin{document}
\title{Quantum discord for two-qubit $X$-states\,: A comprehensive approach inspired by classical polarization optics}
\author{Krishna Kumar Sabapathy}
\email{kkumar@imsc.res.in} 
\affiliation{Optics \& Quantum Information Group,
The Institute of Mathematical
  Sciences, C.I.T Campus, Tharamani, Chennai 600 113, India.}

\author{R. Simon}
\email{simon@imsc.res.in} 
\affiliation{Optics \& Quantum Information Group, The Institute of Mathematical
  Sciences, C.I.T Campus, Tharamani, Chennai 600 113, India.}

\begin{abstract} 
Classical correlation and quantum discord are computed for two-qubit $X$-states. Our approach, which is inspired by the methods of classical polarization optics, is geometric in the sense that the entire analysis is tied to the correlation ellipsoid of all normalized conditional states of the A-qubit with measurement elements applied to the B-qubit. Aspects of the computation which depend on the location of the reduced state of A inside the ellipsoid get clearly separated from those which do not. Our treatment is comprehensive\,: all known results are reproduced, often more economically, and several new insights and results emerge. Detailed reexamination of the famous work of Ali, Rau, and Alber [Phys. Rev. A {\bf 81}, 042105 (2010)], in the light of ours and a counterexample we manufacture against their principal theorem points to an uncommon situation in respect of their principal result\,: their theorem turns out to be `numerically correct' in all but a very tiny region in the space of $X$-states, notwithstanding the fact that their proof of the theorem seems to make, in the disguise of an unusual group theoretic argument, an {\em a priori} assumption equivalent to the theorem itself.
\end{abstract}


\maketitle 

\section{Introduction} 
The study of correlations in bipartite systems has been invigorated over the last couple of decades or so. Various measures and approaches to segregate the classical and quantum contents of correlations have been explored. Entanglement has continued to be the most popular of these correlations owing to its inherent potential advantages in performing quantum computation and communication tasks\,\cite{horo-rmp}.  More recently, however, there has been a rapidly growing interest in the study of correlations from a more direct measurement perspective\,\cite{modi-rmp,celeri-rmp}, and several measures to quantify the same have been considered. Among these measures, quantum discord and classical correlation have been attracting much attention\,\cite{disc-imp1}, and have lead to several interesting results\,\cite{disc-imp2,disc-imp3}. 

In this work, we undertake a comprehensive analysis of the problem of computation of  correlations in the two-qubit system, especially  the so-called $X$-states\,\cite{xstates1}; this class of states has come to be accorded a distinguished status in this regard\,\cite{xstates2}. The problem of $X$-states has already been considered in\,\cite{luo,ali,wang11,chen11,du-geom,du12,huang13} and that of more general two-qubit states in\,\cite{james11,adesso11,zambrini11,zambriniepl,adesso11b,nassajour13}. The approach which we present here fully exploits the very geometric nature of the problem. In addition to being comprehensive, it helps to clarify and correct some  issues in the literature regarding computation of correlations in $X$-states.  It may be emphasised that the geometric methods used here have been the basic tools of (classical) polarization optics for a very long time, and involve elementary constructs like Stokes vectors, Poincar\'{e} sphere, and Mueller matrix\,\cite{simon-mueller1,simon-mueller2,simon-mueller3,simon-mueller4}. 
In this sense our approach to quantum discord and classical correlation is one inspired by classical polarization optics.

We assume, unless otherwise stated, that measurements are performed on subsystem $B$.  The expression for quantum discord ${\cal D}(\hat{\rho}_{AB})$ is then given by\,\cite{zurek01,vedral01}
\begin{align}
{\cal D}(\hat{\rho}_{AB}) &= I(\hat{\rho}_{AB}) - C(\hat{\rho}_{AB}),\nonumber\\
I(\hat{\rho}_{AB}) &= S(\hat{\rho}_A) + S(\hat{\rho}_B) - S(\hat{\rho}_{AB}),
\end{align}
where $I(\hat{\rho}_{AB})$ denotes the mutual information which is supposed to capture the total correlation in the given bipartite state $\hat{\rho}_{AB}$\,\cite{winter05}. The second quantity $C(\hat{\rho}_{AB})$ is the maximum amount of classical correlation that one could extract as a result of  measurements on subsystem $B$.
Now, classical correlation in a bipartite state $\hr{AB}$ is given by the expression\,\cite{vedral01}
\begin{align}
C(\hat{\rho}_{AB}) = \underset{\Pi^B}{{\rm max}} \left[S(\hat{\rho}_A) - \sum_j p_j S(\hat{\rho}^A_j) \right],
\label{cc} 
\end{align} 
where the probabilities $\{ p_j\}$ are given by
\begin{align}
p_j = {\rm Tr}[(1\!\!1_A \otimes \Pi_j^B) \hat{\rho}_{AB}],
\end{align}
 the (normalized) state of system $A$ after measurement $\Pi^B_j$ being given by 
\begin{align}
 \hat{\rho}_j^A = \frac{{\rm Tr}_B[\Pi_j^B \hat{\rho}_{AB}]}{p_j}.
\end{align}
The set $\Pi^B=\{ \Pi^B_j \}$ meets the defining conditions  $\sum_j \Pi^B_j = 1\!\!1$, and $\Pi^B_j \geq 0$ for all $j$. That is,  $\{ \Pi_j^B\}$ forms a POVM. The second term in the expression \eqref{cc} for classical correlation  is the (minimum, average) conditional entropy post measurement, and we may denote it by
\begin{align}
S^A_{\rm min} = \underset{\Pi^B}{{\rm min}} \sum_{j} p_j S(\hat{\rho}^A_j),
\label{samin}
\end{align}  
the minimum being taken over the set of all POVM's. Then the expression for classical correlation simply reads as 
\begin{align}
C(\hat{\rho}_{AB}) =  S(\hat{\rho}_A) - S^A_{\rm min},
\end{align}
and, consequently, that for quantum discord as
\begin{align}
{\cal D}(\hat{\rho}_{AB}) =  S(\hat{\rho}_B) - S(\hat{\rho}_{AB}) + S^A_{\rm min}.
\end{align}
Finally, we note that the first two terms of this expression for quantum discord are known as soon as the bipartite state $\hat{\rho}_{AB}$ is specified. Therefore the only quantity of computational interest is the conditional entropy $S^A_{\rm min}$  of system $A$ post measurement (on B)\,: this  alone involves an optimization. It is to the task of computing $S^A_{\rm min}$ that the methods of classical polarization optics seem to be the most appropriate tools.\\

The content of the paper is organised as follows. We begin by indicating in Section II why we believe that the Mueller-Stokes formalism of classical polarization optics is the most appropriate tool for analyzing conditional states post measurement, and hence for computing quantum discord of a two-qubit state. The Mueller matrix associated with the density operator of an $X$-state is presented in Section III, and the correlation ellipsoid of (normalized) conditional states post measurement associated with a two-qubit state (or its Mueller matrix) is analysed in Section IV. With the geometric tool of the correlation ellipsoid on hand, the problem of computation of the optimal mean conditional entropy $S^A_{\rm min}$ is taken up in Sections V and VI. Our primary aim in Section V is to prove that the present optimization problem is one of convex optimization over an ellipse rather than over an ellipsoid, and hence optimization over a single variable. The actual computation of $S^A_{\rm min}$ is comprehensively treated in Section VI, bringing out clearly all possible situations that could arise. With measurements assumed to be carried out on the B-side, the computational aspects which depend on the reduced state $\hr{A}$ are clearly demarcated from those which do not. The correlation ellipsoid has an invariance group which is much larger than the group of local unitaries. The manner in which this larger invariance group helps the analysis is discussed in Section VII, including the manner in which it helps to connect the separability of a two-qubit $X$-state directly to its correlation ellipsoid. Section VIII is devoted to a detailed comparison of our results with those of Ali, Rau and Alber\,\cite{ali}. $X$-states of vanishing discord are fully enumerated in Section IX, and contrasted with earlier enumerations. Finally, $X$-states for which the discord can be written down by inspection, with no need for optimization, are considered in Section X. This family is much larger than the $X$-states treated in the well-known work of Luo\,\cite{luo}. 

A comment may be in order before we turn to presentation of our analysis and results. While our treatment is geometrical in nature, the emphasis is on comprehensiveness. Thus, while many of our results are new, it is possible that some are known in scattered form in the works of earlier authors. For instance, while it is known from some earlier publications\,\cite{wang11,chen11,adesso11,zambrini11,zambriniepl,du12,huang13} that the main theorem of Ali, Rau, and Alber numerically fails for some $X$-states, the present work seems to be the first to demonstrate that their very proof of the theorem itself is untenable in a fundamental manner. While we prove that the numerical failure applies to only a very tiny region of the space of $X$-states, the failure of their proof would seem to apply not just to this tiny region but to all $X$-states since the symmetry on which they base the proof of their theorem is a property of generic $X$-states. 

\section{Mueller-Stokes formalism for two-qubit states}
We begin with a brief indication as to why the Mueller-Stokes formalism of classical optics is possibly the most appropriate tool for handling quantum states post measurement. In classical polarization optics the state of a light beam is represented by a $2 \times 2$ complex positive matrix $\Phi$ called the {\em polarization matrix}\,\cite{neillbook}. The intensity of the beam is identified with ${\rm Tr}\,\Phi$, and so the  matrix $({\rm Tr}\,\Phi)^{-1} \Phi$ (normalized to unit trace) represents the actual {\em state of polarization}. The polarization matrix $\Phi$ is thus analogous to the density matrix of a qubit, the only distinction being that the trace of the latter needs to assume unit value. Even this one little difference is gone when one deals with {\em conditional quantum states} post measurement\,: the probability of obtaining a conditional state becomes analogous to intensity $={\rm Tr}\, \Phi$ of the classical context. 

The Mueller-Stokes formalism itself arises from the following simple fact\,: any  $2\times 2$ matrix $\Phi$ can be invertibly associated with a  four-vector $S$, called the Stokes vector, through
\begin{align}
\Phi = \frac{1}{2} \sum_{k=0}^3 S_k \sigma_k, ~~ S_k = {\rm Tr}(\sigma_k \Phi).
\end{align} 
This representation is an immediate consequence of the fact that the Pauli triplet $\sigma_1,\,\sigma_2,\,\sigma_3$ and $\sigma_0 = 1\!\!1$, the unit matrix, form a complete orthonormal set of (hermitian) matrices. 

Clearly, hermiticity of the polarization matrix $\Phi$ is {\em equivalent} to reality of the associated four-vector $S$ and ${\rm Tr}\,\Phi = S_0$. Positivity of $\Phi$ reads $S_0 >0$, $S_0^2 -S_1^2-S_2^2-S_3^2 \geq 0$ corresponding, respectively, to the pair ${\rm Tr}\, \Phi>0$, ${\rm det}\, \Phi \geq 0$. Thus positive $2\times 2$ matrices (or their Stokes vectors) are in one-to-one correspondence with points of the {\em positive branch of the solid light cone}. Unit trace (intensity) restriction corresponds to the section of this cone at unity along the `time' axis, $S_0=1$. The resulting three-dimensional unit ball ${\cal B}_3 \in {\cal R}^3$ is the more familiar Bloch (Poincar\'{e}) ball, whose surface or boundary ${\cal P} = {\cal S}^2$ representing pure states (of unit intensity) is often called the  Bloch (Poincar\'{e}) sphere. The interior points correspond to mixed (partially polarized) states. 
 
Optical systems which map Stokes vectors {\em linearly} into Stokes vectors have been of particular interest in polarization optics. Such a linear system is represented by a $4 \times 4$ real matrix $M$, the Mueller matrix\,\cite{simon-mueller1,simon-mueller2,simon-mueller3,simon-mueller4}\,:
\begin{align}
M\,: S^{\rm in} \to S^{\rm out} = M S^{\rm in}.
\label{mspol}
\end{align}
It is evident that a (physical) Mueller matrix should necessarily map the positive solid light cone into itself. {\em It needs to respect an additional subtle restriction}, even in classical optics. \\

\noindent
{\bf Remark 1}\,: The Mueller-Stokes formulation of classical polarization optics traditionally assumes plane waves. It would appear, within such a framework, one need not possibly place on a Mueller matrix any more demand than the requirement that it map Stokes vectors to Stokes vectors. However, the very possibility that the input (classical) light could have its polarization and spatial degrees of freedom intertwined in an inseparable manner, leads to the additional requirement that the Mueller matrix acting `locally' on the polarization indices alone map such an entangled (classical) beam into a physical beam at the output. Interestingly, it is only recently that such an entanglement-based requirement has been established\,\cite{simon-mueller3,simon-mueller4},  leading to a full characterization of Mueller matrices in classical polarization optics.\,$\blacksquare$

To see the connection between Mueller matrices and two-qubit states unfold naturally, use a single index rather than a pair of indices to label the computational basis two-qubit states $\{|jk \rangle\}$ in the familiar manner\,: $(00,01,10,11) = (0,1,2,3)$. Now note that a two-qubit density operator $\hat{\rho}_{AB}$ can be expressed in {\em two distinct ways}\,:
\begin{align}
\hat{\rho}_{AB} &= \sum_{j,k=0}^3 \rho_{jk} |j \rangle \langle k| \nonumber\\
&= \frac{1}{4} \sum_{a,b=0}^3 M_{ab} \,\sigma_a \otimes \sigma_b^*,
\label{mmatrix}
\end{align}
the second expression simply arising from the fact that the sixteen hermitian  matrices $\{\sigma_a \otimes \sigma_b^*  \}$ form a complete orthonormal set of $4 \times 4$ matrices. Hermiticity of operator $\hat{\rho}_{AB}$ is equivalent to reality of the  matrix $M = ((M_{ab}))$, but the same hermiticity is equivalent to  $\rho = ((\rho_{jk}))$ being a hermitian matrix. \\

\noindent
{\bf Remark 2}\,: It is clear from the defining equation\,\eqref{mmatrix} that the numerical entries of the two matrices $\rho,\,M$ thus associated with a given two-qubit state $\hat{\rho}_{AB}$ be related in an invertible linear manner. This linear relationship has been in use in polarization optics for a long time\,\cite{simon-mueller1,simon-mueller4}  and, for convenience, it is reproduced in explicit form in the Appendix.\,$\blacksquare$

Given a bipartite state $\hat{\rho}_{AB}$, the reduced density operators $\hat{\rho}_A,\,\hat{\rho}_{B}$ of the subsystems are readily computed from the associated $M$\,:
\begin{align}
\hat{\rho}_A &= {\rm Tr}[\hat{\rho}_{AB}] = \frac{1}{2} \sum_{a=0}^3 M_{a0}\, \sigma_a, \nonumber\\
\hat{\rho}_B &= {\rm Tr}[\hat{\rho}_{AB}] = \frac{1}{2} \sum_{b=0}^3 M_{0b}\, \sigma_b^*.
\label{partial}
\end{align} 
That is, the leading column and leading row of $M$ are precisely the Stokes vectors of reduced states $\hat{\rho}_A,\,\hat{\rho}_B$ respectively. 

It is clear that a generic  POVM element is of the form 
$\Pi^B_j = \frac{1}{2} \sum_{k=0}^3 S_k \sigma_k^*$. We shall call $S$  the Stokes vector of the POVM element $\Pi^B_j$. Occasionally one finds it convenient to write it in the form $S = (S_0, \mathbf{S})^T$ with the `spatial' $3$-vector part highlighted. 
The Stokes vector corresponding to a rank-one element has components that satisfy the relation $S_1^2 + S_2^2+ S_3^2 =S_0^2$.
Thus, rank-one elements  are  light-like and rank-two elements are strictly time-like. 
One recalls that similar considerations apply to the density operator of a qubit as well. 

The (unnormalised) state operator post measurement (measurement element $\Pi_j$) evaluates to
\begin{align}
\rho_{\pi_j}^A&= {\rm Tr}_B[\hat{\rho}_{AB}\, \Pi_j^B  ] \nonumber\\
&= \frac{1}{8} {\rm Tr}_B\left[\, \left(\sum_{a,b=0}^3 M_{ab}\, \sigma_a \otimes \sigma_b^*\right)\,\left( \sum_{k=0}^3 S_k \sigma^*_k \right)\,\right]\nonumber \\
&=\frac{1}{8}  \sum_{a,b=0}^3 \sum_{k=0}^3 M_{ab} \, S_k \,\sigma_a {\rm Tr}( \sigma_b^* \sigma^*_k )\nonumber\\
&= \frac{1}{4} \sum_{a=0}^3 S^{\,'}_a \sigma_a,
\label{msa}
\end{align}  
where we used  ${\rm Tr}( \sigma_b^* \sigma^*_k ) = 2 \delta_{bk}$ in the last step. \\

\noindent
{\bf Remark 3}\,: It may be emphasised, for clarity, that we use Stokes vectors to represent both measurement elements and states. For instance,  Stokes vector $S$ in Eq.\,\eqref{msa} stands for a measurement element $\Pi^B_j$ on the B-side, whereas $S^{\,'}$ stands for (unnormalised) state of subsystem $A$.\,$\blacksquare$ 

The Stokes vector of the resultant state in Eq.\,\eqref{msa} is thus given by 
 $S^{\,'}_a= \sum_{k=0}^3 M_{ak} S_k$, which may be written in the suggestive form
\begin{align}
S^{\rm out} = M S^{\rm in}.
\label{ms}
\end{align}
Comparison with \eqref{mspol} prompts one to call $M$ the {\em Mueller matrix associated with two-qubit state} $\hat{\rho}_{AB}$. 
We repeat that the conditional state $\rho_{\pi_j}^A$ need not have unit trace, and so needs to be normalised when computing entropy post measurement. To this end, we write 
\begin{align}
\rho_{\pi_j}^A &= p_{j} \hat{\rho}_{\pi_j}\nonumber\\
 p_{j}   = \frac{S^{\rm out}_0}{2}, ~~ \hat{\rho}_{\pi_j} &= \frac{1}{2} (1\!\!1 + (S_0^{\rm out})^{-1} \,\mathbf{S}^{\rm out}.\boldsymbol{\sigma}).
\end{align}
It is sometimes convenient to write the Mueller matrix $M$ associated with a given state $\hat{\rho}_{AB}$  in the block form 
\begin{align}
M = \left( 
\begin{array}{cc}
1 & \boldsymbol{\xi}^T \nonumber\\
\boldsymbol{\lambda} & \Gamma
\label{genm}
\end{array}
\right), ~~ \boldsymbol{\lambda},\,\boldsymbol{\xi} \in {\cal R}^3.
\end{align}
Then the input-output relation \eqref{ms} reads 
\begin{align}
S_{0}^{\rm out} = S_0^{\rm in} + \boldsymbol{\xi} \cdot \mathbf{S^{\rm in}}, ~~~ \mathbf{S^{\rm out}} &= S_0^{\rm in}\,\boldsymbol{\lambda} + \Gamma\, \mathbf{S^{\rm in}},
\end{align}
showing in particular that the probability of the conditional state $S^{\rm out}$ on the A-side depends on the POVM element precisely through $\boldsymbol{\xi}\cdot {\bf S}^{\rm in}$.\\

\noindent
{\bf Remark 4}\,: The linear relationship between two-qubit density operators $\rho$ (states) and Mueller matrices (single qubit maps) we have developed in this Section can be usefully viewed as an instance of the Choi-Jamiokowski isomorphism\,\cite{cj}.\,$\blacksquare$\\

\noindent
{\bf Remark 5}\,: We have chosen measurements to be made on the B qubit. Had we instead chosen to compute correlations by performing measurements on subsystem $A$ then, by similar considerations as detailed above, we would have found $M^T$ playing the role of the Mueller matrix $M$.\,$\blacksquare$ 


\section{$X$-states and their Mueller matrices}
$X$-states are states whose density matrix $\rho$ has non-vanishing entries only along  the diagonal and the anti-diagonal. That is, the numerical  matrix $\rho$ has the `shape' of $X$. A general $X$-state can thus be written, to begin with, as
\begin{align}
\rho_X = \left( \begin{array}{cccc}
\rho_{00} & 0 &0 & \rho_{03} e^{i \phi_2} \\
0& \rho_{11} & \rho_{12} e^{i \phi_1} &0 \\
0&\rho_{12} e^{-i\phi_1} & \rho_{22} &0\\ 
\rho_{03} e^{-i \phi_2} & 0 & 0& \rho_{33}
\end{array}
\right),
\end{align}
where the $\rho_{ij}$'s are all real nonnegative.  One can get rid of the phases (of the off-diagonal elements) by a suitable local unitary transformation
$U_A \otimes U_B$. This is not only possible, but {\em also  desirable} because the quantities of interest, namely entanglement, mutual information, quantum discord and classical correlation, are all invariant under local unitary transformations. Since it is unlikely to be profitable to carry around a baggage of irrelevant parameters, we shall indeed shed $\phi_1,\,\phi_2$ by taking $\rho_X$ to its canonical form $\rho_X^{\rm can}$.
We have  
\begin{align}
\rho_X \to \rho_X^{\rm can} =U_A \otimes U_B \, \rho_X \, U_A^{\dagger} \otimes U_B^{\dagger},
\end{align}
 where
\begin{align}
\rho_X^{\rm can}  &= \left( \begin{array}{cccc}
\rho_{00} & 0 &0 & \rho_{03}  \\ 
0& \rho_{11} & \rho_{12}  &0 \\
0&\rho_{12} & \rho_{22} &0\\ 
\rho_{03}  & 0 & 0& \rho_{33}
\end{array}
\right);\nonumber\\
U_A &= {\rm diag}(e^{-i(2\phi_1+ \phi_2)/4},e^{i \phi_2/4}),\nonumber\\ U_B &= {\rm diag}(e^{i( 2\phi_1-\phi_2)/4},e^{i \phi_2/4}).
\label{xcan}
\end{align}

\noindent
{\bf Remark 6}\,: We wish to clarify that $X$-states thus constitute, in the canonical form, a (real) 5-parameter family, three diagonal parameters ($\rho_{00}+\rho_{11}+\rho_{22}+\rho_{33} = m_{00}=1$) and two off-diagonal parameters;  it can be lifted,  using  local unitaries $U_A,\,U_B \in SU(2)$ which have three parameters each, to a $11$-parameter subset in the $15$-parameter state space (or generalized Bloch sphere) of two-qubit states\,: they are all local unitary equivalent to the conventional $X$-states, though they may no more have  `shape' $X$.\,$\blacksquare$  

With this canonical form, it is clear that the Mueller matrix for the generic $X$-state $\rho_X^{\rm can}$  has the form
\begin{align}
M = \left( \begin{array}{cccc}
1 & 0&0&m_{03}\\
0&m_{11}&0&0\\
0&0&m_{22} &0\\
m_{30}&0&0&m_{33}
\end{array}\right), 
\label{mueller}
\end{align} 
where 
\begin{align}
m_{11} &= 2(\rho_{03} + \rho_{12}), ~~m_{22} =  2(\rho_{03} - \rho_{12}), \nonumber \\
m_{03} &= \rho_{00} + \rho_{22} - (\rho_{11} + \rho_{33}), \nonumber \\
m_{33} &= \rho_{00} + \rho_{33} - (\rho_{11} + \rho_{22}), \nonumber\\
m_{30} &= \rho_{00} + \rho_{11} - (\rho_{22} + \rho_{33}),
\end{align}
as can be read off from the defining equation \eqref{mmatrix} or from the relation in the Appendix. We note that the Mueller matrix of an $X$-state has a `sub-X' form\,: the only nonvanishing off-diagonal entries are $m_{03}$ and $m_{30}$ ($m_{12} = 0 = m_{21}$). In our computation later we will sometimes need the inverse relations
\begin{align}
\rho_{00} &= \frac{1}{4} (m_{00}+m_{03}+m_{30}+m_{33}),\nonumber\\\rho_{11}&=\frac{1}{4} (m_{00}-m_{03}+m_{30}-m_{33}),\nonumber\\ 
\rho_{22} &= \frac{1}{4} (m_{00}+m_{03}-m_{30}-m_{33}),\nonumber\\\rho_{33}&=\frac{1}{4} (m_{00}-m_{03}-m_{30}+m_{33}),\nonumber\\
\rho_{03} &= \frac{1}{4} (m_{11}+m_{22}),~~\rho_{12}=\frac{1}{4} (m_{11}-m_{22}).
\end{align}

The positivity properties of $\rho^{\rm can}_X$, namely $\rho_{00}\, \rho_{33} \geq \rho_{03}^2$, $\rho_{11}\, \rho_{22} \geq \rho_{12}^2$  transcribes to the following conditions on the entries of its Mueller matrix\,:
\begin{align}
(1+m_{33})^2 &- (m_{30}+m_{03})^2 \geq (m_{11}+m_{22})^2 \label{cp1}\\
(1-m_{33})^2 &- (m_{30}-m_{03})^2 \geq (m_{11}-m_{22})^2.
\label{cp2}
\end{align}

\noindent
{\bf Remark 7}\,: As noted earlier the requirements \eqref{cp1}, \eqref{cp2} on  Mueller matrix \eqref{mueller} in the classical polarization optics context was established for the first time in Refs.\,\cite{simon-mueller3,simon-mueller4}. These correspond to complete positivity requirement on $M$ considered as a positive map (map which images the solid light cone into itself), and turns out to be equivalent to positivity of the corresponding two-qubit density operator.\,$\blacksquare$

By virtue of the direct-sum block structure of $X$-state density matrix, one can readily write down its (real) eigenvectors. We choose the following order for definiteness\,:
\begin{align}
|\psi_0 \rangle &=\, c_{\alpha} |00\rangle + s_{\alpha} |11\rangle,~~|\psi_1 \rangle = \,c_{\beta} |01\rangle + s_{\beta} |10\rangle, \nonumber\\
|\psi_2 \rangle &= -s_{\beta} |01\rangle + c_{\beta} |10\rangle, \, |\psi_3 \rangle = -s_{\alpha}  |00\rangle + c_{\alpha}  |11\rangle,
\label{spec1}
\end{align}
where $c_{\alpha},\,s_{\alpha}$ denote respectively $\cos{\alpha}$ and $\sin{\alpha}$. And (dropping the superscript `can') we have the spectral resolution  
\begin{align}
\hat{\rho}_X = \sum_{j=0}^3 \lambda_j |\psi_j \rangle \langle \psi_j|,
\label{spec2}
\end{align} 
\begin{align}
c_{\alpha} &= \sqrt{\frac{1+\nu_1}{2}}, ~ c_{\beta} = \sqrt{\frac{1+\nu_2}{2}},\nonumber\\
\nu_1 &= \frac{\rho_{00}-\rho_{33}}{\sqrt{4\rho_{03}^2 +( \rho_{00} - \rho_{33})^2}} = \frac{m_{30}+m_{03}}{\sqrt{(m_{11}+m_{22})^2 +(m_{30}+m_{03})^2}},\nonumber\\
\nu_2 &= \frac{\rho_{11}-\rho_{22}}{\sqrt{4\rho_{12}^2 +( \rho_{11} - \rho_{22})^2}} = \frac{m_{30}-m_{03}}{\sqrt{(m_{11}-m_{22})^2 +(m_{30}-m_{03})^2}};\nonumber\\
\lambda_{0 \,{\rm or}\, 3} &= \frac{\rho_{00}+\rho_{33}}{2} \pm \frac{\sqrt{(\rho_{00}-\rho_{33})^2 + 4\,\rho_{03}^2}}{2}\nonumber\\
&= \frac{1+m_{33}}{4} \pm \frac{\sqrt{(m_{11}+m_{22})^2 +(m_{30}+m_{03})^2}}{4},\nonumber\\
\lambda_{1\, {\rm or}\, 2} &= \frac{\rho_{11}+\rho_{22}}{2} \pm \frac{\sqrt{(\rho_{11}-\rho_{22})^2 + 4\,\rho_{12}^2}}{2}\nonumber\\
&= \frac{1-m_{33}}{4} \pm \frac{\sqrt{(m_{11}-m_{22})^2 +(m_{30}-m_{03})^2}}{4}.
\label{eigen}
\end{align}
 
While computation of $S^A_{\rm min}$ will have to wait for a detailed consideration of the manifold of conditional states of $\hat{\rho}_{AB}$, the other entropic quantities can be evaluated right away. Given a qubit state specified by Stokes vector $(1,\mathbf{S})^T$, it is clear that its von Neumann entropy equals
\begin{align}
S_2(r)=-\left[\frac{1+r}{2}\right]\, {\ell og}_2 {\left[\frac{1+r}{2}\right]} -\left[\frac{1-r}{2}\right]\, {\ell og}_2 {\left[\frac{1-r}{2}\right]},
\end{align}
where $r$ is the norm of the three vector $\mathbf{S}$, or the distance of $\mathbf{S}$ from the origin of the Bloch ball.
Thus from Eq.\,\eqref{partial} we have
\begin{align}
S(\hat{\rho}_A) &= S_2(|m_{30}|), ~S(\hat{\rho}_B)=S_2(|m_{03}|), \nonumber\\
S(\hat{\rho}_{AB}) &\equiv S_2(\{\lambda_j \})= \sum_{j=0}^3 -\lambda_j {\ell og}_2\,(\lambda_j), 
\label{ents}
\end{align}
where $\lambda_j$, $j=0,1,2,3$ are the eigenvalues of the bipartite state $\hat{\rho}_{AB}$ given in Eq.\,\eqref{eigen}. The mutual information thus assumes the value
\begin{align}
I(\hat{\rho}_{AB}) = S_2(|m_{30}|) + S_2(|m_{03}|) - S_2(\{\lambda_j \}).
\label{mi}
\end{align}

\section{Correlation ellipsoid\,: Manifold of conditional states}
We have seen that the state of subsystem $A$ resulting from measurement of any POVM element on the B-side of $\hat{\rho}_{AB}$ is the Stokes vector resulting from
the action of the associated Mueller matrix on the Stokes vector of the POVM element. 
In the case of rank-one measurement elements, the `input' Stokes vectors correspond to light-like points on the (surface ${\cal S}^2 = {\cal P}$  of the) Bloch ball. Denoting the POVM elements as
$S^{\rm in}=(1,x,y,z)^T$, $x^2+y^2+z^2=1$, we ask for the collection of corresponding normalized  conditional states. By Eq.\,\eqref{ms} we have 
\begin{align}
S^{\rm out} &= M S^{\rm in} = \left( \begin{array}{c} 1+m_{03} z\\ m_{11} \,x \\ m_{22} \, y\\ m_{30} + m_{33} \,z \end{array}\right) 
\to \left( \begin{array}{c} 1 \\ \frac{m_{11} \,x}{1+m_{03} z} \\ \frac{m_{22} \, y}{1+m_{03} z}\\ \frac{m_{30} + m_{33} \,z}{1+m_{03} z} \end{array}\right).
\label{outstates}
\end{align}
It is clear that, for $S_0^{\rm in}=1$,  $S_0^{\rm out} \neq 1$ whenever $m_{03} \neq 0$ and the input is {\em not} in the x-y plane of the Poincar\'{e} sphere.
It can be shown that the sphere $x^2+y^2+z^2=1$ at the `input' is mapped to the ellipsoid 
\begin{align}
\frac{x^2}{a_x^2} + \frac{y^2}{a_y^2} +\frac{(z-z_c)^2}{a_z^2} = 1
\end{align}
of {\em normalized states} at the output, the parameters of the ellipsoid being fully determined by the entries of $M$\,: 
\begin{align}
a_x &= \frac{|m_{11}|}{\sqrt{1-m_{03}^2}}, ~~~ a_y= \frac{|m_{22}|}{\sqrt{1-m_{03}^2}}, \nonumber \\ 
a_z &= \frac{|m_{33} - m_{03}m_{30}|}{1-m_{03}^2}, ~~~z_c = \frac{m_{30} - m_{03} m_{33}}{1-m_{03}^2}.
\label{ellipsoid}
\end{align}

\noindent
{\bf Remark 8}\,: This ellipsoid of all possible (normalized) conditional states associated with a two-qubit state is sometimes known as the steering ellipsoid\,\cite{du-geom,du12,rudolph13}. It degenerates into a single point if and only if the state is a product or uncorrelated state. It captures in a geometric manner correlations in the two-qubit state under consideration, and correlation is the object of focus in the present work. For these reasons, we prefer to call it the {\em correlation ellipsoid} associated with the given two-qubit state. While measurement elements $\Pi_j$ are mapped to points of the ellipsoid, measurement elements $a \Pi_j$ for all $a>0$ and fixed $\Pi_j$ are mapped to one and the same point of the correlation ellipsoid. Thus, in the general case, each point of the ellipsoid corresponds to a `ray' of measurement elements. In the degenerate case wherein the ellipsoid becomes a disc or line segment or a single point and only in that case, do several rays map to the same point.\,$\blacksquare$

\begin{figure}
\begin{center}
\scalebox{0.8}{\includegraphics{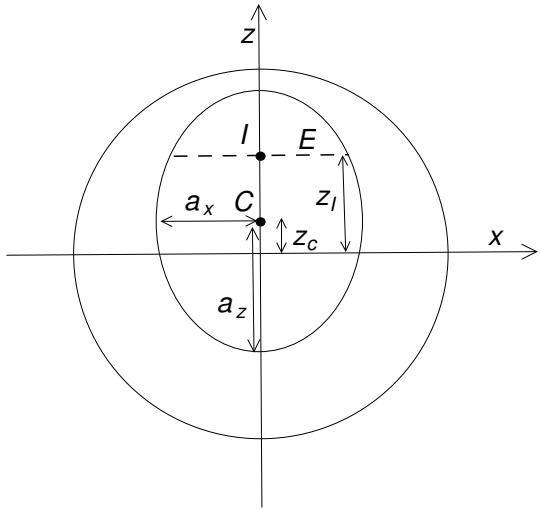}}
\end{center}
\caption{Showing the x-z section of the correlation ellipsoid associated with a generic $X$-state. The point ${\rm I}$ represents the location of $\hr{A}$, the image of the maximally mixed input, $C$ the center of the ellipsoid, and $E$ represents the image of the equatorial plane of the input Bloch sphere. 
 \label{ellipsoida}} 
\end{figure}

The x-z section of the correlation ellipsoid is pictorially depicted in Fig.\,\ref{ellipsoida}. It is clear that the geometry of the ellipsoid is determined by the four parameters $a_x, a_y,a_z,z_c$ and $z_c$ could be assumed nonnegative without loss of generality. The fifth parameter $m_{30}$ specifying the z-coordinate of the image ${\rm I}$ of the maximally mixed state as measurement element on the B side, is not part of this geometry.  It is clear that ${\rm I}$ corresponds to $\hr{A}$.

Having thus considered the passage from a two-qubit $X$-state to its correlation ellipsoid, we may raise the converse issue of going from the correlation ellipsoid to the associated $X$-state. To do this, however, we need the parameter $z_I=m_{30}$ as an input in addition to the ellipsoid itself. Further, change of the signature of $m_{22}$ does not affect the ellipsoid in any manner, but changes the states and correspondingly the signature of ${\rm det} M$. Thus, the signature of ${\rm det} M$ needs to be recorded as an additional binary parameter. It can be easily seen that the nonnegative $a_x,a_y,a_z,z_c$ along with $z_I$ and ${\rm sgn}\, ({\rm det} M)$ fully reconstruct the $X$-state in its canonical form \eqref{xcan}, \eqref{mueller} [see Remark 16]. Using local unitary freedom we can render $m_{11},\,m_{33}-m_{03} m_{30}$ and $z_c$ nonnegative so that ${\rm sgn}(m_{22}) = {\rm sgn}\, ({\rm det} M)$; $z_I = m_{30}$ can assume either signature. It turns out to be convenient to denote by $\Omega^+$ the collection of all Mueller matrices with ${\rm det} M \geq 0$ and by $\Omega^-$ those with ${\rm det} M \leq 0$. The intersection $\Omega^+ \bigcap \Omega^-$ corresponds to Mueller matrices for which ${\rm det} M=0$, a measure zero subset. Further, in our analysis to follow we assume, without loss of generality,
\begin{align}
a_x \geq a_y,~ i.e.,~ m_{11} \geq |m_{22}|.
\label{axgeqay}
\end{align}

\noindent
{\bf Remark 9}\,: Every two-qubit state has associated with it a unique correlation ellipsoid of (normalized) conditional states. An ellipsoid centered at the origin needs six parameters for its description\,: three for the sizes of the principal axes and three for the orientation of the ellipsoid as a rigid body in ${\cal R}^3$. For a generic (i.e., not necessarily $X$) state , the centre $C$ can be shifted from the origin to vectorial location $\vec{r}_c$, thus accounting for three parameters, and ${\rm I}$ can be located at $\vec{r}_I$ anywhere inside the ellipsoid, thus accounting for another three. The three-parameter local {\em unitary} freedom on the B-side, which has no effect whatsoever on the geometry of the ellipsoid (but determines which points of the input Poincar\'{e} sphere go to which points on the surface of the ellipsoid) accounts for the final three parameters, adding to a total of $15$. For $X$-states the shift of $C$ from the origin {\em needs to be} along one of the principal directions and ${\rm I}$ is {\em constrained to be located on this very principal axis}. In other words, $\vec{r}_c$ and $\vec{r}_I$ become one-dimensional rather than three-dimensional variables rendering $X$-states a 11-parameter subfamily of the 15-parameter state space. {\em Thus $X$-states are distinguished by the fact that $C$, ${\rm I}$, and the origin are collinear with one of the principal axes of the ellipsoid}. This geometric rendering pays no special respect to the shape $X$, but is manifestly invariant under local unitaries as against the characterization in terms of `shape' $X$ of the matrix $\rho_{AB}$ in the computation basis. {\em Since the latter (conventional) characterization is not even invariant under local unitaries, we are tempted to a strong appeal to the community in favour of our invariant geometric characterization of $X$-states.}\,$\blacksquare$

\section{Optimal measurement \label{optsec}}
In this Section we take up the central part of the present work which is to develop a provably optimal scheme for computation of the quantum discord for any $X$-state of a two-qubit system. Our treatment  is {\em both comprehensive and self-contained and, moreover, it is geometric in flavour}. We begin by exploiting symmetry to show, without loss of generality, that the problem itself is one of {\em optimization in just a single variable}. The analysis is entirely based on the output or correlation ellipsoid associated with a two-qubit state $\hr{AB}$, and we continue to assume that measurements are carried out on the B-side. 

The single-variable function under reference will be seen, on optimization, to divide the manifold of possible correlation ellipsoids into two subfamilies. For one subfamily the optimal measurement or POVM will be shown to be a von Neumann measurement along either x or z, {\em independent of the location (inside the ellipsoid) of ${\rm I}$, the image of the maximally mixed input}. For the other subfamily, the optimal POVM will turn out to be either a von Neumann measurement along x or a three-element POVM, {\em depending on the actual location of ${\rm I}$ in the ellipsoid}. There exists no $X$-state for which the optimal measurement requires a four-element POVM, neither does there exist an $X$-state for which the optimal POVM is von Neumann in a direction which is neither along x nor z.

For the special case of the centre $C$ of the ellipsoid coinciding with the origin $z=0$ of the Poincar\'{e} sphere ($z_c=0$), it will be shown that the optimal measurement is always a von Neumann measurement along x or z, {\em irrespective of the location of $z_I$ in the ellipsoid}. While this result may look analogous to the simple case of Bell mixtures earlier treated by Luo\,\cite{luo}, it should be borne in mind that these centred $X$-states form a much larger family than the family of Bell mixtures, for in the Luo scenario {\em ${\rm I}$ necessarily coincides with $C$ and  with the origin}, but we place no such restriction of coincidence. Stated differently, in our case of centered ellipsoids  $z_I$ is an independent variable in addition to $a_x,a_y,a_z$. We shall return to the case of centered ellipsoids in Section X. 

As we now turn to the analysis itself it is useful to record this\,: the popular result that the optimal POVM requires no more than four elements plays {\em a priori} no particular role of help in our analysis; it is for this reason that we shall have no occasion in our analysis to appeal to this important theorem\,\cite{zaraket04,ariano05}. \\

\noindent
{\bf Proposition 1}\,: The optimal POVM needs to comprise  rank-one elements.\\
\noindent
{\em Proof}\,: This fact is nearly obvious, and equally obvious is its proof. Suppose $\omega_j$ is a rank-two element of an optimal POVM and $\hat{\rho}_j^A$ the associated conditional state of subsystem $A$. Write $\omega_j$ as a positive (convex) sum of rank-one elements $\omega_{j1},\,\omega_{j2}$ and let $\hat{\rho}_{j1}^A,\,\hat{\rho}_{j2}^A$ be the conditional states corresponding respectively to $\omega_{j1},\,\omega_{j2}$.  It is then clear that $\hat{\rho}_j = \lambda \hat{\rho}_{j1}^A + (1-\lambda) \hat{\rho}_{j2}^A$, for some $0 < \lambda <1$. Concavity of the entropy function $S$ immediately implies $S(\hat{\rho}_j^A) > \lambda S(\hat{\rho}_{j1}^A) + (1-\lambda) S(\hat{\rho}_{j2}^A)$, in turn implying through \eqref{samin} that the POVM under consideration could not have been optimal, thus completing the proof. It is clear from the nature of the proof that this fact applies to all states and not just to $X$-states, and to all Hilbert space dimensions and not just $d=2$.\,$\blacksquare$\\

\noindent 
{\bf Remark 10}\,: Since a rank-one POVM element $|v \rangle \langle v|$ is just a point $\mathbf{S}$ on (the surface of) the Bloch (Poincar\'{e}) sphere ${\cal P}$, a four element rank-one POVM is a quadruple of points $\mathbf{S^{(j)}}$ on ${\cal P}$, with associated probabilities $p_j$. The POVM condition $\sum_j p_j |v_j \rangle \langle v_j| = 1\!\!1$ demands that we have to solve the pair
\begin{align}
p_1 + p_3 &+p_3 +p_4 =2,\nonumber\\
&\sum_j p_j\,\mathbf{S^{(j)}} =0.
\label{povmcondition}
\end{align}
Once four points $\mathbf{S^{(j)}}$ on ${\cal P}$ are chosen, the `probabilities' $\{p_j \}$ are {\em not independent}. To see this, consider the tetrahedron for which  $\mathbf{S^{(j)}}$ are the vertices. If this tetrahedron does not contain the origin, then $\sum_j p_j \mathbf{S^{(j)}} =0$ has no solution with nonnegative $\{p_j \}$. If it contains the origin, then there exits a solution and the solution is `essentially' unique by Caratheodory theorem. 

The condition $\sum_j p_j =2$ comes into play in the following manner. Suppose we have a solution to $\sum_j p_j\,\mathbf{S^{(j)}} =0$. It is clear that $p_j \to p_j^{\,'} = a p_j$, $j=1,2,3,4$, with no change in $\mathbf{S^{(j)}}$'s, will also be a solution for any ($j$-independent) $a>0$. It is this freedom in choosing the scale parameter $a$ that gets frozen by the condition $\sum_j p_j =2$, rendering the association between tetrahedra and solutions of the pair \eqref{povmcondition} indeed unique.

{\em We thus arrive at a geometric understanding of the manifold of all (rank-one) four-element POVM's (even though one would need such POVM's only when one goes beyond $X$-states). This is precisely the manifold of all tetrahedra with vertices on ${\cal P}$, and containing the centre in the interior of ${\cal P}$.} We are not considering four-element POVM's whose $\mathbf{S^{(j)}}$ are coplanar with the origin of ${\cal P}$, because they are of no use as optimal measurements. It is clear that three element rank-one POVM's  are similarly characterized, again by the Caratheodory theorem, by triplets of points on ${\cal P}$ coplanar with the origin of ${\cal P}$, with the requirement that the triangle generated by the triplet contains the origin {\em in the interior}. Further, it is trivially seen in this manner that  2-element rank-one POVM's are von Neumann measurements determined by pairs of antipodal $\mathbf{S^{(j)}}$'s on ${\cal P}$, i.e., by `diameters' of ${\cal P}$.\,$\blacksquare$

The correlation ellipsoid of an $X$-state (as a subset of the Poincar\'{e} sphere) has a ${\cal Z}_2 \times {\cal Z}_2$ symmetry generated by reflections respectively about the x-z and y-z planes. We shall now use the product of these two reflections---a $\pi$-rotation or inversion about the z-axis---to simplify, without loss of generality, our problem of optimization. \\

\noindent
{\bf Proposition 2}\,: All elements of the optimal POVM have to necessarily correspond to (light-like) Stokes vectors of the form $S_0(1,\, \sin{\theta},\,0,\,\cos{\theta})^T$, i.e., the measurement elements are constrained to the x-z plane. \\
\noindent
{\em Proof}\,: Suppose ${\cal N} = \{\omega_1,\,\omega_2,\,\cdots,\,\omega_k \}$ is an optimal POVM of rank-one elements (we are placing no restriction on the cardinality $k$ of ${\cal N}$, but rather expect it to unfold naturally from the analysis to follow). And let $\{\hat{\rho}_1^A,\,\hat{\rho}_2^A,\,\cdots,\,\hat{\rho}_k^A \}$ be the corresponding (normalized) conditional states, these being points on the boundary of the correlation ellipsoid. Let $\tilde{\omega}_j$ and $\tilde{\hat{\rho}}_j^A$ represent, respectively, the images of $\omega_j$, $\hat{\rho}_j^A$ under $\pi$-rotation about the z-axis (of the input Poincar\'{e} sphere and of the correlation ellipsoid)\,: $\tilde{\omega}_j = \sigma_3 \, \omega_j\,\sigma_3$, $\tilde{\hat{\rho}}_j^A = \sigma_3\,\hr{j}^A\,\sigma_3$. It follows from symmetry that $\widetilde{{\cal N}} = \{\tilde{\omega}_1,\,\tilde{\omega}_2,\,\cdots,\,\tilde{\omega}_k \}$ too is an optimal POVM. And so is also ${\cal N}\, \overline{\bigcup} \,\widetilde{{\cal N}}$, where we have used the decorated symbol $\overline{\bigcup}$ rather than the set union symbol $\bigcup$ to distinguish from simple union of sets\,: if $S_0(1\!\!1 \pm \sigma_3)$ happens to be an element $\omega_j$ of ${\cal N}$, then $\tilde{\omega}_j = \omega_j$ for this element, and in that case this $\omega_j$ should be `included' in ${\cal N}\, \overline{\bigcup}\, \widetilde{{\cal N}}$ {\em not once but twice} (equivalently its `weight' $S_0$ needs to be doubled). The same consideration holds if ${\cal N}$ includes any $\omega_j$ and $\sigma_3 \, \omega_j\,\sigma_3$.

Our supposed to be optimal POVM can thus be assumed to comprise pairs of elements $\omega_j,\,\tilde{\omega}_j$ related by inversion about the z-axis. Let us consider the associated pair of conditional states $\hat{\rho}_j^A,\,\tilde{\hat{\rho}}_j^A$ on the (surface of the) correlation ellipsoid. They have identical z-coordinate $z_j$. The section of the ellipsoid (parallel to the x-y or equatorial plane) at $z=z_j$ is an ellipse, with major axis along $x$ (recall \eqref{axgeqay} wherein we have assumed, without loss of generality, $a_x \geq a_y$), and $\hat{\rho}_j^A$ and $\tilde{\hat{\rho}}_j^A$ are on opposite ends of a line segment through the centre $z_j$ of the ellipse. Let us {\em assume} that this line segment is not the major axis of the ellipse $z=z_j$. That is, we assume $\hat{\rho}_j^A$, $\tilde{\hat{\rho}}_j^A$ are not in the x-z plane. 

Now slide (only) this pair along the ellipse smoothly, keeping them at equal and opposite distance from the z-axis until both reach opposite ends of the major axis of the ellipse, the x-z plane. It is clear that during this process of sliding both $\hat{\rho}_j^A$, $\tilde{\hat{\rho}}_j^A$ recede away from the centre of the ellipse and hence away from the centre of the Poincar\'{e} sphere itself. As a result $S(\hat{\rho}_j^A)$ decreases, thus improving the value of $S^A_{\rm min}$ in \eqref{samin}. This would have proved that the POVM ${\cal N}$ is not optimal, unless our  assumption that $\hat{\rho}_j^A$, $\tilde{\hat{\rho}}_j^A$ are not in the x-z plane is false. This completes proof of the proposition.\,$\blacksquare$

This preparation immediately leads to the following important result which forms the basis for our further analysis.

\begin{theorem}\,:
The problem of computing quantum discord for $X$-states is a problem of convex optimization on a plane, and optimization over a single variable.   
\end{theorem}
\noindent
{\em Proof}\,: We have just proved that elements of the optimal POVM come, in view of the ${\cal Z}_z \times {\cal Z}_2$ symmetry of $X$-states,  in pairs $S_0(1,\,\pm \sin{\theta},\,0,\,\cos{\theta})^T$ of Stokes vectors $\omega_j,\,\tilde{\omega}_j$ with $0 \leq \theta \leq \pi$. The corresponding conditional states come in pairs $\hat{\rho}_j^A$, $\tilde{\hat{\rho}}_j^A = 1/2(1\!\!1 \pm x_j \sigma_1 + z_j \sigma_3)$. The two states of such a pair of conditional states are at the same distance 
\begin{align}
r(z_j) = \sqrt{z_j^2 + a_x ^2 -(z_j -z_c)^2 a_x^2/a_z^2}
\label{rz}
\end{align}
from the origin of the Poincar\'{e} sphere, and hence they have the same von Neumann entropy 
\begin{align}
f(z_j) &= S_2(r(z_j)),\nonumber\\
S_2(r)&= -\frac{1+r}{2}\,{\ell og}_2\,\frac{1+r}{2}-\frac{1-r}{2}\,{\ell og}_2\,\frac{1-r}{2}.
\label{fz}
\end{align}

Further, continuing to assume without loss of generality $a_x \geq a_y$, our convex optimization is not over the three-dimensional ellipsoid, but effectively a {\em planar} problem over the x-z elliptic section of the correlation ellipsoid (Proposition 2), and hence the optimal POVM cannot have more that three elements. Thus, the (Stokes vectors of the) optimal POVM elements on the B-side necessarily have the form,
\begin{align}
\Pi^{(3)}_{\theta} &= \{ 2 p_0(\theta) (1,0,0,1)^T, \,2 p_1(\theta) (1, \pm \sin{\theta}, 0, -\cos{\theta} )^T\}, \nonumber\\
 p_0(\theta) &= \frac{\cos{\theta}}{1+ \cos{\theta}}, ~~ p_1(\theta) = \frac{1}{\cos{\theta}}, ~ 0 \leq \theta \leq \pi/2.
\label{scheme}
\end{align}
The optimization itself is thus over the single variable $\theta$.\,$\blacksquare$\\

\noindent
{\bf Remark 11}\,: It is clear that $\theta=0$ and $\theta = \pi/2$ correspond respectively to von Neumann measurement along z and x, and {\em no other von Neumann measurement gets included} in $\Pi^{(3)}_{\theta}$. Every $\Pi^{(3)}_{\theta}$ in the open interval $0 < \theta < \pi/2$ corresponds to a {\em genuine} three-element POVM. The symmetry considerations above do allow also three-element POVM's of the form 
\begin{align}
\widetilde{\Pi}^{(3)}_{\theta} &= \{ 2 p_0(\theta) (1,0,0,-1)^T,\, 2 p_1(\theta) (1, \pm \sin{\theta}, 0, \cos{\theta} )^T\}, ~~
 0 \leq \theta \leq \pi/2,
\end{align} 
but such POVM's lead to local maximum rather than  minimum for $S^A$, and hence are of no value to us.\,$\blacksquare$

\section{Computation of $S^A_{\rm min}$ \label{saminsec}}
\begin{figure}
\begin{center}
\scalebox{0.9}{\includegraphics{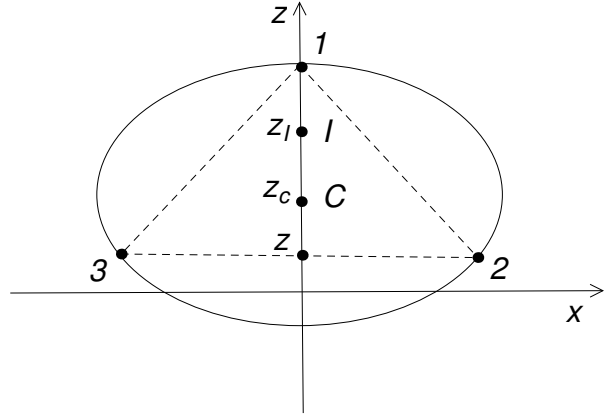}}
\end{center}
\caption{Showing the conditional states corresponding to the 3-element measurement scheme of\,\eqref{scheme}. The points 1,2,3 on the surface of the correlation ellipsoid represent the conditional states corresponding to the three measurement elements of \eqref{scheme}.} 
\label{opta}
\end{figure}

A schematic diagram of the 3-element  POVM $\Pi^{(3)}_{\theta}$ of Eq.\,\eqref{scheme} is shown in Fig.\,\ref{opta}. The Bloch vectors of the corresponding conditional states $\hr{1}^A,\, \hr{2}^A,\, \hr{3}^A$ at the output are found to be of the form 
\begin{align}
(0,\,0,\,z_c+a_z)^T,\,&(x(z),\,0,\,z)^T,\,(-x(z),\,0,\,z)^T, \nonumber\\
~~ &x(z) = \frac{a_x}{a_z}(a_z^2 - (z-z_c)^2)^{1/2}.
\end{align}
For these states denoted $1,2,3$ in Fig.\,\ref{opta} the weights should be chosen  to realize as convex sum the state ${\rm I}$ (the image of the maximally mixed input) whose Bloch vector is  $(0,0,z_I)^T$.  von Neumann measurements along the z or x-axis correspond respectively to $z= z_c-a_z$ or $z = z_I$. 
Using Eqs.\,\eqref{rz},\,\eqref{fz}, the expression for $S^A(z)$ is thus given by 
\begin{align}
S^A(z)= p_1(z) \, f(z_c+a_z) + p_2(z)\,f(z),\nonumber\\
p_1(z) = \frac{z_I-z}{z_c+a_z-z}, ~~ p_2(z) = \frac{z_c+a_z-z_I}{z_c+a_z-z}. 
\label{actual}
\end{align}
The minimization of $S^A(z)$ with respect to the single variable $z$ should give $S^A_{\rm min}$. It may be noted in passing that, for a given ${\rm I}$ or $z_I$, the three-element POVM parametrized by $z$ makes no sense in the present context for $z > z_I$ (since $p_1(z)$ ought to be $\geq 0$).

\begin{figure}
\begin{center}
\scalebox{0.8}{\includegraphics{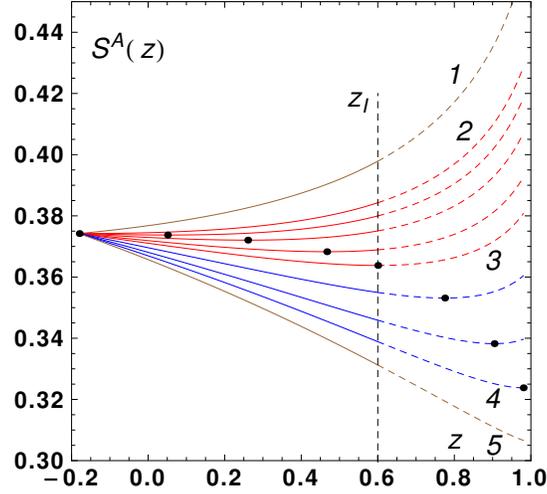}}
\end{center}
\caption{Showing $S^A(z)$ for various values of $a_x$, the curves being labelled in increasing order of $a_x$. The vertical line marked $z_I$ denotes the reference $z=z_I$. A three-element POVM scheme could potentially result for values of $a_x \in (a_x^V(a_z,z_c), a^H_x(a_z,z_c))$ [the region between curves (2) and (4)]. For values of $a_x \leq a^V(a_z,z_c)$ [curve (2) and above],  von Neumann projection along the z-axis is the optimal POVM and for values of $a_x \geq a^H(a_z,z_c)$ [curve (4) and below],  von Neumann projection along the x-axis is the optimal one. The optimal $z_0$ (marked by a dot) for a curve is obtained by minimizing $S^A(z)$ on the curve. The function $S^A(z)$ in \eqref{actual} is not meaningful for $z>z_I$, and this region is demarcated by the reference vertical line at $z=z_I$ and distinguished with dashed curves. In this example $(a_z,z_c) = (0.58,0.4)$. For $z_I = 0.6$, a three-element POVM results for all (red) curves between (2) and (3). For this value of $z_I$, curves (3) and below correspond to horizontal von Neumann projection being the optimal POVM.} 
\label{sap}
\end{figure}

For clarity of presentation, we begin by considering a specific example  $(a_z,\,z_c,\,z_{I}) = (0.58,\,0.4,0.6)$. To begin with, the relevant interval for the variable $z$ in this case is $[z_c-a_z,\, z_c+a_z]=[-0.18,0.98]$, and we shall examine the situation as we vary $a_x$ for fixed $(a_z,z_c,z_{I})$. The behaviour of $S^A(z)$ for this example is depicted in Fig.\,\ref{sap}, wherein each curve in the $(z,\,S^A(z))$ plane corresponds to a chosen value of $a_x$, and the value of $a_x$ increases as we go down Fig.\,\ref{sap}. For values of $a_x \leq a_x^V(a_z,z_c)$, for some $a_x^V(a_z,z_c)$ to be detailed later, $S^A(z)$ is seen to be a monotone increasing function of $z$, and so its minimum $S^A_{\rm min}$ obtains at the `lower' end point $z = z_c-a_z=-0.18$, hence the optimal POVM corresponds to the vertical projection or von Neumann measurement along the z-axis. The curve marked $2$ corresponds to $a_x = a^V_x(a_z,z_c)$ [which equals $0.641441$ for our example].

Similarly for values of $a_x \geq a_x^H(a_z,z_c)$, $S^A(z)$ proves to be a monotone decreasing function of $z$, its minimum therefore obtains at the `upper' end point which is $z_I$ and not $z_c+a_z$ [recall that the three-element POVM makes no sense for $z>z_I$]; hence the optimal POVM corresponds to horizontal projection or von Neumann measurement along x-axis. The curve marked 4 corresponds to $a_x = a^H_x(a_z,z_c)$ [which equals $0.677305$ for our example]. It will be shown later that both $a^V_x(a_z,z_c)$, $a^H_x(a_z,z_c)$ do indeed depend only on $a_z,z_c$ and not on $z_I$. Both are therefore properties of the ellipsoid\,: all states with one and the same ellipsoid share the same $a^V_x(a_z,z_c)$, $a^H_x(a_z,z_c)$.

Thus, it is the region $a^V_x(a_z,z_c) < a_x < a^H_x(a_z,z_c)$ of values of $a_x$ that needs a more careful analysis, for it is only in this region that the optimal measurement {\em could possibly correspond} to a three-element POVM. Clearly, this region in the space of correlation ellipsoids is distinguished by the fact that $S^A(z)$ has a minimum at some value $z = z_0$ in the open interval $(z_c-a_z,\,z_c+a_z)$. For $a_x = a^V_x(a_z,z_c)$ this minimum occurs at $z_0 = z_c- a_z$, moves with increasing values of $a_x$ gradually towards $z_c+a_z$, and reaches $z_c+a_z$ itself as $a_x$ reaches $a^H_x(a_z,z_c)$. 

\begin{figure}
\begin{center}
\scalebox{0.8}{\includegraphics{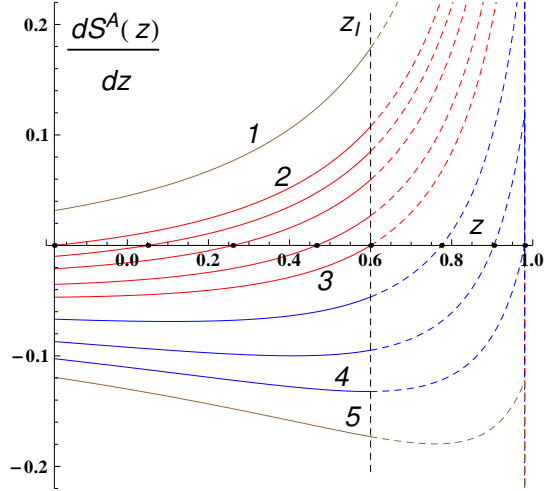}}
\end{center}
\caption{Showing $dS^A(z)/dz$ for various values of $a_x$, the curves being labelled in increasing order of $a_x$. A root $z_0$ exits  for values of $a_x \in (a_x^V(a_z,z_c), a^H_x(a_z,z_c))$ [between curves (2) and (4)]. For values of $a_x \leq a^V(a_z,z_c)$ and  $a_x \geq a^H(a_z,z_c)$, there is no root $z_0$ [curves (1) and (5) being examples].} 
\label{firstderi}
\end{figure}

Not only this qualitative behaviour, but also the exact value of $z_0(a_z,z_c,a_x)$ is independent of $z_I$. Let us evaluate $z_0(a_z,z_c,a_x)$ by looking for the zero-crossing of the derivative function $dS^A(z)/dz$ depicted in Fig.\,\ref{firstderi}. We have 
\begin{align}
\frac{d\, S^A(z)}{dz} &= (z_c+a_z-z_I)\,G(a_z,a_x,z_c;z),\nonumber\\
G(a_z,a_x,z_c;z) &= \frac{1}{(z_c+a_z-z)^2}\left([(z_c+a_z-z)(a_x^2(z-z_c)/a_z^2-z) X(z)]\right.\nonumber\\
 &~~~~~~~~~~~~~~~~~~~~~~~~~~~~~~~~~~~~ \left. - [f(z_c+a_z)-f(z)]\right),\nonumber\\
X(z) &= \frac{1}{2r(z)} {\ell og}_2\left[\frac{1+r(z)}{1-r(z)}\right],
\label{Gexp}
\end{align} 
and we need to look for $z_0$ that solves $G(a_z,a_x,z_c;z_0)=0$. The reader may note that $z_I$ {\em does not enter} the function $G(a_z,a_x,z_c;z_0)$ defined in Eq.\,\eqref{Gexp}, showing that $z_0$ is indeed independent of $z_I$ as claimed earlier\,: {\em $z_0$ is a property of the correlation ellipsoid; all states with the same correlation ellipsoid have the same $z_0$}.

Let us focus on the two curves $a^V_x(a_z,z_c)$, $a^H_x(a_z,z_c)$ alluded to earlier and defined through 
\begin{align}
a^V_x(a_z,z_c)\,:~  G(a_z,a_x^V,z_c;z_c-a_z) =0,\nonumber\\
a^H_x(a_z,z_c)\,:~ G(a_z,a_x^H,z_c;z_c+a_z) =0.
\label{bounds}
\end{align}
The  curve $a^V_x(a_z,z_c)$ characterizes, for a given $(a_z,z_c)$, the value of $a_x$ for which the first derivative of $S^A(z)$ vanishes at $z= z_c-a_z$ (i.e., $z_0 = z_c - a_z$), so that the vertical von Neumann projection is the optimal POVM for all $a_x \leq a^V_x(a_z,z_c)$. Similarly, the  curve $a^H_x(a_z,z_c)$ captures the value of $a_x$ for which the first derivative of $S^A(z)$ vanishes at $z= z_c+a_z$.  
Solving for the two curves in terms of $a_z$ and $z_c$ we obtain, after some algebra, 
\begin{align}
a^V_x(a_z,z_c) &= \sqrt{\frac{f(z_c-a_z)-f(z_c+a_z)}{2 X(z_c-a_z)} - a_z(z_c-a_z)},\nonumber\\
a^H_x(a_z,z_c) &= \frac{(z_c+a_z)}{2[Y(z_c+a_z)-X(z_c+a_z)]}\nonumber\\
& \hspace{2cm}\times  \, \left[(z_c-a_z) X(z_c+a_z) + 2a_zY(z_c+a_z) - \sqrt{w}\,\right],\nonumber\\
Y(z) &= \frac{1}{[{\ell n}\,2\,](1-r(z)^2)},\nonumber\\
~~w &= X(z_c+a_z)[(z_c-a_z)^2 X(z_c+a_z) + 4 a_z z_c Y(z_c+a_z)].
\label{ahav}
\end{align}
\begin{figure}
\begin{center}
\scalebox{0.8}{\includegraphics{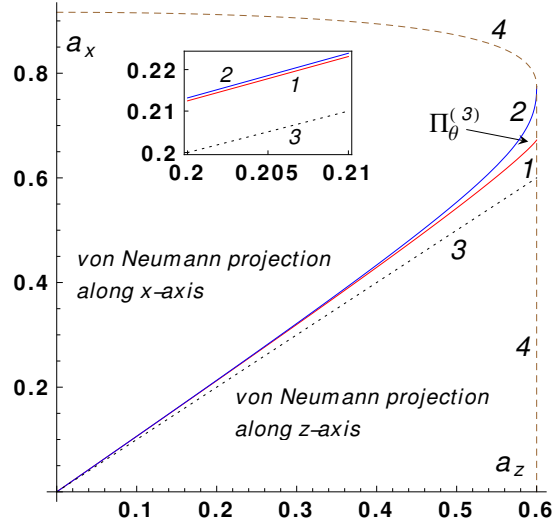}}
\end{center}
\caption{Showing the various possibilities for optimal measurement scheme across a slice (the $a_x-a_z$ plane) of the  parameter space (of correlation ellipsoids) with $z_c$ fixed at $z_c=0.4$.  Only for a tiny wedge-shaped region marked $\Pi^{(3)}_{\theta}$, the region between $a^V(a_z,z_c)$ [curve (1)] and $a^H(a_z,z_c)$ [curve (2)],  can one expect a {\em potential} 3-element POVM. For the region above (2), von Neumann measurement along the x-axis proves optimal and for region below curve (1)  von Neumann measurement along the z-axis is optimal. Curve marked (4) depicts the boundary of allowed values for $a_z,a_x$ (CP condition). The curve (3) is the line $a_z = a_x$. The inset resolves curves (1), (2) and (3) in the small region $a_x \in [0.2,0.21]$, to emphasize that these curves remain distinct (except at $a_x=a_z=0$).} 
\label{wedge}
\end{figure}
These curves are marked $(1)$ and $(2)$ respectively  in Fig.\,\ref{wedge}. Two aspects are of particular importance\,:
\begin{enumerate}[(i)]
\item $a_x^H(a_z,z_c) \geq a^V_x(a_z,z_c)$, the inequality saturating if and only if $z_c=0$. In particular these two curves never meet (except at $a_x=a_z=0$), the appearance in Fig.\,\ref{wedge} notwithstanding. It is to emphasize this fact that an inset has been added to this figure. The straight line $a_x = a_z$, marked (3) in Fig.\,\ref{wedge}, shows that $a_x^V(a_z,z_c) \geq  a_z$, the inequality again saturating if and only if $z_c=0$.
\item It is only in the range $a_x^V(a_z,z_c) < a_x < a_x^H(a_z,z_c)$ that we get a solution $z_0$
\end{enumerate}
\begin{align}
G(a_z,a_x,z_c;z_0) =0, ~~ z_c-a_z < z_0 < z_c+ a_z
\end{align}
corresponding to a {\em potential} three-element optimal POVM for {\em some} $X$-state corresponding to the ellipsoid under consideration; and, clearly, the optimal measurement for the state will {\em actually} correspond to a three-element POVM only if 
\begin{align}
a_z + z_c > z_I > z_0. 
\label{threerange}
\end{align}
If $z_I \leq z_0$, the optimal measurement corresponds to a von Neumann projection along the x-axis, and never the z-axis.

\begin{figure}
\begin{center}
\scalebox{0.8}{\includegraphics{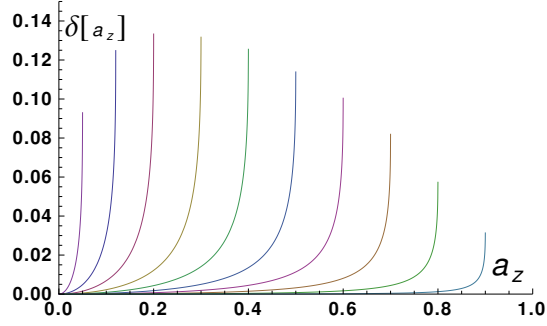}}
\end{center}
\caption{Showing $\delta(a_z)$ for various values of $z_c$, decreasing from left to right. The first (left most) curve corresponds to $z_c=0.95$ and the last one to $z_c=0.1$. The size of the `wedge'-shaped region of Fig.\,\ref{wedge} is seen to first increase and then decrease with increasing $z_c$.} 
\label{shifts}
\end{figure}

We  note that the range of values of $a_x$ for a fixed $(a_z,z_c)$ where a potential three-element POVM can exist, the width of the `wedge' region in Fig.\,\ref{wedge}, increases with increasing $a_z$. 
Let us define a parameter $\delta$ through $\delta(a_z,z_c) = a^H_x(a_z,z_c) - a^V_x(a_z,z_c)$ to capture the extent of the region bounded by curves (1) and (2) in Fig.\,\ref{wedge}. This object is shown in Fig.\,\ref{shifts}. 
We see that the range of values of $a_x$ for which a three-element POVM exists first increases with increasing $z_c$ and then decreases.\\

\noindent
{\bf Remark 12}\,:
In the special case $z_c=0$ of centered correlation ellipsoids it is clear from \eqref{ahav} that curves (1) and (2) of Fig.\,\ref{wedge} coincide with the linear curve $a_x=a_z$ marked (3). And this behaviour is independent of the value of $-a_z < z_I < a_z$. As a consequence, the optimal POVM for centered ellipsoids is a z or x von Neumann projection according as $a_x  < a_z$ or $a_x > a_z$. We shall return to this fact in Section X.\,$\blacksquare$\\

\begin{figure}
\begin{center}
\scalebox{0.8}{\includegraphics{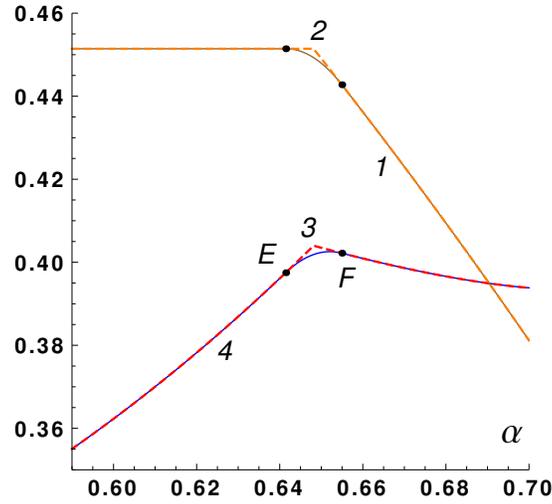}}
\end{center}
\caption{
Showing $S^A_{\rm min}$ [curve (1)] and quantum discord [curve (4)] for a one-parameter family of states $\hr{}(\alpha)$ with parameters $(a_x,a_y,a_z,z_c) = (\alpha,0.59,0.58,0.4)$ and  $z_I=0.5$, as $\alpha$ varies over the range $[0.59,0.7]$. We have $a^V_x(0.58,0.4)=0.641441$ and $a^H_x(0.58,0.4)=0.677305$. Point E ($a_x(E)=0.641441$) denotes the transition of the optimal measurement from a von Neumann measurement along the z-axis to a three-element POVM, while point F ($a_x(F) = 0.654947$) denotes the transition of the optimal measurement from a three-element POVM to a von Neumann measurement along the x-axis. The curve (3) (or (2))  denotes the over-estimation of quantum discord (or $S^A_{\rm min}$) by restricting the measurement scheme to a von Neumann measurement along the z or x-axis. } 
\label{example1}
\end{figure}

\noindent
{\bf \em An illustrative example}\,:
We now evaluate quantum discord for a one-parameter family of states which we denote by $\hr{}(\alpha)$. The Mueller matrix associated with $\hr{}(\alpha)$ is chosen as\,:
\begin{align}
M(\alpha) = \left[\begin{array}{cccc}
1 & 0 & 0 & y\\
0& \alpha (1-y^2)^{1/2} & 0 & 0 \\
0& 0 & 0.59 (1-y^2)^{1/2} & 0\\
0.5 & 0 & 0 & 0.58 + 0.4 y
\end{array}
\right],
\end{align}
with $y=0.1/0.58$ and $\alpha \in [0.59,0.7]$. The ellipsoid parameters for our class of states are given by $(a_x,a_y,a_z,z_c) = (\alpha,0.59,0.58,0.4)$, and $z_I=0.5$. These states are in $\Omega^+$, and differ from one another only in  $a_x$ which changes as the parameter $\alpha$ is varied in the  interval $[0.59,0.7]$. Using the optimal measurement schemes outlined above and in the earlier Section, we compute $S^A_{\rm min}$ and the quantum discord. The results are depicted in Fig.\,\ref{example1}, with $S^A_{\rm min}$  denoted by the solid curve (1) and quantum discord by the solid curve (4).  `Over-estimation' of quantum discord by restricting to von Neumann measurements along x or z-axis is shown by the broken curve (3) for comparison with the optimal three-element POVM curve (4), and this corresponds to the corresponding over-estimation of $S^A_{\rm min}$ shown as broken curve (2). The point E denotes the transition from z-axis projection to a three-element POVM and point F denotes the transition from the three-element POVM to  von Neumann measurement along the x-axis.

\begin{figure}
\scalebox{0.8}{\includegraphics{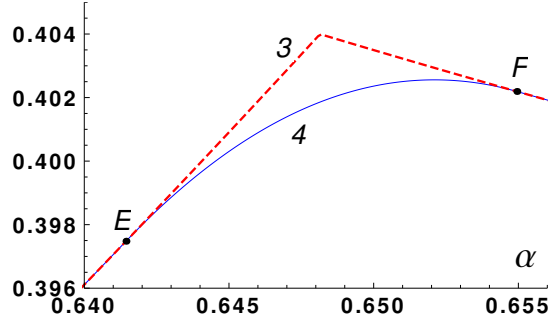}}
\caption{
Showing the region between E and F of Fig.\,\ref{example1} in amplified form. What would have been paraded as a sharp transition in the Ali, Rau, and Alber\,\cite{ali} scheme [curve (3)] is clearly seen, in reality, to be a smooth affair [curve (4)]. 
\label{example1b}}
\end{figure}

For purpose of clarity, the region of Fig.\,\ref{example1} between E and F is amplified in Fig.\,\ref{example1b}. What would have appeared as a sharp transition had one stuck to von Neumann projections  along x and z (the Ali, Rau, Alber prescription) is clearly seen, in reality, to be a smooth affair.  \\


\noindent
{\bf \em  Purification and  EoF}\,:
The Koashi-Winter theorem or relation\,\cite{koashi04} shows that the classical correlation  of a given bipartite state $\hat{\rho}_{AB}$ is related to the entanglement of formation of the `complimentary' state $\hat{\rho}_{CA}$. 
That is, 
\begin{align}
C(\hat{\rho}_{AB}) = S(\hat{\rho}_A) - E_F(\hat{\rho}_{CA}).
\end{align}
Comparing with the definition of $S^A_{\rm min}$ in Eq.\,\eqref{samin}, we see that
\begin{align}
 S^A_{\rm min}(\hat{\rho}_{AB}) = E_F(\hat{\rho}_{CA}).
\end{align}
In other words, the Koashi-Winter relation connects and equates the (minimum average) conditional entropy post measurement of a bipartite state $\hat{\rho}_{AB}$ to the 
entanglement of formation of its complimentary state $\hat{\rho}_{CA}$, the latter state being defined through purification of $\hat{\rho}_{AB}$ to pure state $|\phi_{C:AB}\rangle$.

The purification can be written as $|\phi_{C:AB} \rangle = \sum_{j=0}^3 \sqrt{\lambda_j}  |e_j \rangle \otimes |\psi_j\rangle$, $\{|e_j\rangle\}$ being orthonormal vectors in the Hilbert space of subsystem C. And the complimentary state $\hat{\rho}_{CA}$ results when subsystem $B$ is simply discarded\,:
\begin{align}
\hat{\rho}_{CA} &= {\rm Tr}_B [|\phi_{C:AB}\rangle \langle \phi_{C:AB}|] \nonumber\\
& = \sum_{j,k=0}^3 \sqrt{\lambda_j \lambda_k} |e_j \rangle \langle e_k| \otimes {\rm Tr}[|\psi_j\rangle \langle \psi_k|].
\end{align}
It is easy to see that for the case of the two qubit $X$-states, the complimentary state belongs to a $2 \times 4$ system. Now that $S^A_{\rm min}$ is determined for all $X$-states by our procedure, using Eqs.\,\eqref{spec1},\,\eqref{spec2},\,\eqref{eigen}  one can immediately write down the expressions for the entanglement of formation for the complimentary states of the $2\times 4$ system that correspond to the entire 11-parameter family of $X$-states. This would not, of course, determine the entanglement of formation of all states of a $2\times 4$ system (the state space being a $63$-parameter convex set), but only for a $11$-parameter subfamily thereof. We note in passing that examples of this connection for  particular cases of states such as rank-two two-qubit states and Bell-mixtures have been earlier studied in\,\cite{du-rank2,yan11}. Thus what we have here is a generalization of these studies.

\section{Invariance group beyond local unitaries}
Recall that a measurement element (on the B side) need not be normalized. Thus in constructing the correlation ellipsoid associated with a two-qubit state $\hr{AB}$, we gave as input to the Mueller matrix associated with $\hr{AB}$ an arbitrary four-vector in the positive solid light cone (corresponding to an arbitrary $2 \times 2$ positive matrix), and then normalized the output Stokes vector to obtain the image point on the correlation ellipsoid. It follows, on the one hand, that all measurement elements which differ from one another by positive multiplicative factors lead to the same image point on the correlation ellipsoid. On the other hand it follows that  $a\,\hr{AB}$ has the same correlation ellipsoid as $\hr{AB}$, for all $a>0$. As one consequence, it is not essential to normalize a  Mueller matrix to $m_{00}=1$ as far as construction of the correlation ellipsoid is concerned. 

The fact that construction of the correlation ellipsoid deploys the entire positive solid light cone of positive operators readily implies that {\em the ellipsoid inherits all the symmetries of this solid light cone}. These symmetries are easily enumerated. Denoting by $\psi_1,\,\psi_2$ the computational basis components of a vector $|\psi\rangle$ in Bob's Hilbert space ${\cal H}_B$, a nonsingular linear transformation 
\begin{align}
J\,: \begin{pmatrix} \psi_1 \\ \psi_2 \end{pmatrix} \to \begin{pmatrix} \psi_1^{\,'} \\ \psi_2^{\,'} \end{pmatrix}
= J \begin{pmatrix} \psi_1 \\ \psi_2 \end{pmatrix}
\end{align}
(or filtering) on ${\cal H}_B$ corresponds on Stokes vectors to the transformation $|{\rm det} J|\,L$ where $L$ is an element of the Lorentz group $SO(3,1)$, and the factor $|{\rm det} J|$ corresponds to `radial' scaling of the light cone\,\cite{simon-mueller1,simon-mueller2}. Following the convention of classical polarization optics, we may call $J$ the {\em Jones matrix} of the (non-singular) local filtering\,\cite{simon-mueller1,simon-mueller2,simon-mueller3,simon-mueller4}. When $({\rm det} J)^{-1/2} J \sim L$ is polar decomposed, the positive definite factor corresponds to pure boosts of $SO(3,1)$ while the (local) unitary factor corresponds to the `spatial' rotation subgroup $SO(3)$ of $SO(3,1)$\,\cite{simon-mueller1,simon-mueller3,simon-mueller4}. It follows that restriction of attention to the section $S_0=1$ (Bloch ball) confines the invariance group from $SO(3,1)$ to $SO(3)$. 

The positive light cone is mapped onto itself also under inversion of all `spatial' coordinates\,: $(S_0, {\bf S}) \to (S_0, -{\bf S})$. This symmetry corresponds to the Mueller matrix $T={\rm diag}(1,-1,-1,-1)$, which is equivalent to $T_0={\rm diag}(1,1,-1,1)$, and hence corresponds to the transpose map on $2\times 2$ matrices. In contradistinction to $SO(3,1)$, $T_0$ acts directly on the operators and cannot be realized or lifted as filtering on Hilbert space vectors; indeed, it cannot be realized as any physical process. Even so, it remains a symmetry of the positive light cone and hence of the correlation ellipsoid itself. 

The full invariance group ${\cal G}$ of a correlation ellipsoid thus comprises two copies of the Lorentz group and the one-parameter semigroup of radial scaling by factor $a>0$\,:
\begin{align}
{\cal G} = \{ SO(3,1), \, T_0 SO(3,1) \approx SO(3,1)T_0,\,a\}.
\label{largesym}
\end{align} 
{\em All Mueller matrices $MM_0$ with $M_0 \in {\cal G}$ and fixed $M$ correspond to one and the same correlation ellipsoid}. In what follows we examine briefly the manner in which these invariances could be exploited for our purpose, and we begin with $SO(3,1)$.

The Jones matrix $J= \exp[\mu \sigma_3/2] = {\rm diag}(e^{\mu/2},e^{-\mu/2})$ corresponds to the Lorentz boost 
\begin{align}
M_0(\mu) = c_{\mu} \left[ 
\begin{array}{cccc}
1 & 0 & 0 & t_{\mu}\\
0& (c_{\mu})^{-1} & 0 & 0\\
0&0&(c_{\mu})^{-1}&0\\
t_{\mu} & 0&0&1
\end{array}
\right]
\label{mboost}
\end{align} 
along the third `spatial' direction on Stokes vectors. Here $c_{\mu},\,t_{\mu}$ stand respectively for $\cosh{\mu}$ and ${\rm tanh}\,{\mu}$. To see the effect of this boost on the correlation ellipsoid, consider a Mueller matrix of the form\,\eqref{mueller} with $m_{03}=0$ so that $m_{11} =a_x$, $m_{33} =a_z$, $m_{22} = \pm a_y$ and $z_c=m_{30}$. Absorbing the scale factor $c_{\mu}$ in Eq.\,\eqref{mboost} into the solid light cone, we have 
\begin{align}
M M_0(\mu) =  
\left[ 
\begin{array}{cccc}
1 & 0 & 0 & t_{\mu}\\
0&  m_{11}/c_{\mu} & 0 & 0\\
0&0&m_{22}/c_{\mu}&0\\
m_{30} + m_{33} \,t_{\mu} & 0&0&m_{33} + m_{30} \,t_{\mu}
\end{array}
\right].
\end{align} 
With the help of \eqref{ellipsoid} we immediately verify that $a_x,\,a_y,\,a_z,$ and $z_c$ associated with $MM_0(\mu)$ are exactly those associated with $M$ with no change whatsoever, consistent with the fact that we expect $M$ and $MM_0(\mu)$ to have the same correlation ellipsoid. Only $z_I$, the image of identity, changes from $m_{30}$ to $m_{30} + m_{33}\,t_{\mu}$\,: as $t_{\mu}$ varies over the permitted open interval $(-1,1)$, the point $z_I$ varies linearly over the open interval $(z_c-a_z, \,z_c+a_z)$. Thus, {\em it is the Lorentz boost on the B side which connects states having  one and the same correlation ellipsoid, but different values of $z_I$}. 

\begin{figure}
\begin{center}
\scalebox{0.8}{\includegraphics{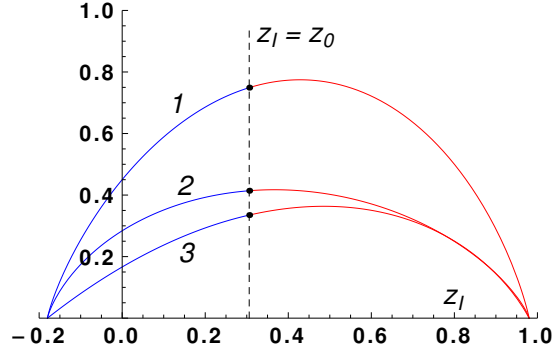}}
\end{center}
\caption{Showing mutual information (1), quantum discord (2), and classical correlation (3) as a function of $z_I$ for fixed ellipsoid parameters $(a_z,z_c,a_x,a_y)=(0.58,0.4,0.65,0.59)$, and hence fixed $z_0=0.305919$. For $z_I \leq  z_0$, the optimal measurement is  von Neumann projection along the x-axis, and for $z_I>z_0$ the optimal measurement is a three-element POVM. The range for $z_I$ is $(z_c -a_z, z_c +a_z) = (-0.18,0.98)$. It should be appreciated that the entire situation corresponds to just one point in the interior of the wedge-shaped region of Fig.\,\ref{wedge}.} 
\label{boost}
\end{figure}

As an illustration of this connection, we go back to Fig.\,\ref{wedge} and consider a correlation ellipsoid corresponding to the point $(a_z,z_c,a_x,a_y)=(0.58,0.4,0.65,0.59)$ in the interior of the wedge region between curves (1) and (2) of Fig.\,\ref{wedge}. We recall that a point in this region is distinguished by the fact that for states corresponding to this point the optimal POVM could {\em potentially} be a three-element POVM, but whether a three element POVM or a horizontal projection actually turns out to be the optimal one for a particular state requires the value of $z_I$ as additional information on the state, beyond the correlation ellipsoid. The behaviour of classical correlation, quantum discord, and mutual information as the Lorentz boost on the B side sweeps  $z_I$ across the full interval $(z_c-a_z,z_c+a_z)$ is presented in Fig.\,\ref{boost}. We repeat that the entire Fig.\,\ref{boost} corresponds to one fixed point in Fig.\,\ref{wedge}.\\

\noindent
{\bf Remark 13}\,: Any entangled two-qubit pure state can be written as 
\begin{align}
|\psi\rangle_{AB} = (1\!\!1 \otimes J)\,|\psi_{\rm max}\rangle,
\end{align}
where the Jones matrix $J$ is non-singular and $|\psi_{\rm max}\rangle$ is a Bell state. Since the associated $SO(3,1)$ does not affect the correlation ellipsoid, the ellipsoid corresponding to $|\psi\rangle_{AB}$ is the same as that of the Bell state, and thereby it is the full Bloch sphere. Hence, $S^A_{\rm min}$ trivially evaluates to zero. Thus we see that for all two-qubit pure states $I(\hr{AB})=2C(\hr{AB})=2D(\hr{AB})=2E(\hr{AB})$.\,$\blacksquare$ \\

\noindent
{\bf Remark 14}\,: It is useful to make two minor observations before we leave the present discussion of the role of $SO(3,1)$. First, it is obvious that a bipartite operator $\hr{AB}$ is positive if and only if its image under any (nonsingular) local filtering $J$ is positive. This, combined with the fact that the location of $z_I$ inside the correlation ellipsoid can be freely moved around using local filtering, implies that the location of $z_I$ has no role to play in the characterization of positivity of $\hr{AB}$ given in \eqref{cp1}, \eqref{cp2}. Consequently, in forcing these positivity requirements on the correlation ellipsoid we are free to move, without loss of generality, to the simplest case corresponding to $z_I = z_c$ or $m_{03}=0$. 

Secondly, since determinant of an $SO(3,1)$ matrix is positive, we see that local filtering does not affect the signature of ${\rm det} M$,
and hence it leaves unaffected the signature of the correlation ellipsoid itself\,: $\Omega^+$ and $\Omega^-$ remain separately invariant under $SO(3,1)$.\,$\blacksquare$

The case of the spatial inversion $T$, to which we now turn our attention, will prove to be quite different on both counts. 
It is clear that the effect of $T_0\,: M \to M T_0$ on an $X$-state Mueller matrix (in the canonical form) is to transform $m_{22}$ to $-m_{22}$, leaving all other entries of $M$ invariant. Since the only way $m_{22}$ enters the correlation ellipsoid parameters in \eqref{ellipsoid} is through $a_y=|m_{22}|$, it follows that the correlation ellipsoid itself is left invariant, but its signature gets reversed\,: ${\rm det}\, MT_0=-\,{\rm det} M$. This reversal of signature of the ellipsoid has important consequences. 

As explained earlier during our discussion of the role of $SO(3,1)$ we may assume, without loss of generality, $z_I = z_c$ or, equivalently, $m_{03}=0$. The positivity conditions \eqref{cp1}, \eqref{cp2} then read as the following requirements on the ellipsoid parameters\,:
\begin{align}
(1+a_z)^2 - z_c^2 \geq (a_x + a_y)^2, \label{x1}\\
(1-a_z)^2 - z_c^2 \geq (a_x - a_y)^2 \label{x2},
\end{align}
in the case $M \in \Omega^+$ (${\rm det}M>0$), and 
\begin{align}
(1+a_z)^2 - z_c^2 \geq (a_x - a_y)^2, \label{x3}\\
(1-a_z)^2 - z_c^2 \geq (a_x + a_y)^2 \label{x4},
\end{align}
in the case $M \in \Omega^-$ (${\rm det}M <0$). But \eqref{x3} is manifestly weaker than \eqref{x4} and hence is of little consequence. The demand that $M T_0$ too correspond to a physical state requires 
\begin{align}
(1+a_z)^2 - z_c^2 \geq (a_x - a_y)^2, \label{x5}\\
(1-a_z)^2 - z_c^2 \geq (a_x + a_y)^2 \label{x6},
\end{align}
in the case $M \in \Omega^+$, and  
\begin{align}
(1+a_z)^2 - z_c^2 \geq (a_x + a_y)^2, \label{x7}\\
(1-a_z)^2 - z_c^2 \geq (a_x - a_y)^2 \label{x8},
\end{align}
in the case of $M \in \Omega^-$.

Now, in the case of $M \in \Omega^+$, \eqref{x5} is weaker than \eqref{x1} and hence is of no consequence, but \eqref{x6} is stronger than and subsumes {\em both} \eqref{x1} and \eqref{x2}. In the case $M \in \Omega^-$ on the other hand both \eqref{x7} and \eqref{x8} are weaker than \eqref{x4}. These considerations establish the following\,:
\begin{enumerate}[1.]
\item  If $M \in \Omega^-$, its positivity requirement is governed by the single condition \eqref{x4} and, further, $M T_0$ certainly corresponds to a physical state in $\Omega^+$.
\item If $M \in \Omega^+$, then $MT_0 \in \Omega^-$ is physical if and only if the additional condition \eqref{x6} which is the same as \eqref{x4} is met.
\end{enumerate}
Since $T_0$ is the same as  partial transpose on the B side, we conclude that a correlation ellipsoid corresponds to a separable state if and only if \eqref{x4} is met, and it may be emphasised that this statement is independent of the signature of the ellipsoid. Stated differently, all $M$ matrices in $\Omega^-$ correspond to separable states, and those in $\Omega^+$ whose signature reversed version are not present in $\Omega^-$ correspond to entangled states. In other words, the set of entangled $X$-states constitute precisely the `$\Omega^-$ complement' of $\Omega^+$.

Finally, this necessary and sufficient condition $(1-a_z)^2 - z_c^2 \geq (a_x + a_y)^2$ for separability can be used to ask for the correlation ellipsoid of maximum volume that corresponds to a separable state, for a given $z_c$. In the case $z_c=0$, it is easily seen that the maximum volume obtains for $a_x = a_y =a_z = 1/3$, and evaluates to a fraction $1/27$ of the volume of the Bloch ball. For $z_c \neq 0$, this fractional volume $V(z_c)$ can be shown to be
\begin{align}
V(z_c) = \frac{1}{54} (2-\sqrt{1+3z_c^2})^2(1+\sqrt{1+3z_c^2}),
\end{align}
and corresponds to 
\begin{align}
a_x = a_y = \frac{[(2-\sqrt{1+3z_c^2})(1+\sqrt{1+3z_c^2})]^{1/2}}{3\sqrt{2}},~~ a_z =\frac{2-\sqrt{1+3z_c^2}}{3}.
\end{align}
 It is a monotone decreasing function of $z_c$. Thus $\Omega^-$ has no ellipsoid of fractional volume $> 1/27$.\\

 \noindent
{\bf Remark 15}\,: It is clear that any $X$-state whose ellipsoid degenerates into an elliptic disc necessarily corresponds to a separable state. This sufficient separability condition may be contrasted with the case of discord wherein the ellipsoid has to necessarily become doubly degenerate into a line segment for nullity of quantum discord to obtain.\,$\blacksquare$  \\

\noindent
{\bf Remark 16}\,: In Section IV we pointed to the fact that a {\em generic} $X$-state is completely specified by the ellipsoid parameters $a_x,a_y,a_z,z_c$ along with $z_I$, the location of the reduced state $\hr{A}$ in the interior of the ellipsoid, and the binary $\epsilon=\pm 1$ distinguishing $\Omega^+$ from $\Omega^-$. We can now present the explicit form for the density matrix $\rho(a_x,a_y,a_z,z_c,z_I,\epsilon)$ associated with the correlation ellipsoid. We have
{\scriptsize \begin{align}
&\rho(a_x,a_y,a_z,z_c,z_I,\epsilon)=\nonumber\\
&\frac{1}{4a_z^4} \left[ 
\begin{array}{cccc}
(1+z_c+a_z)(a_z + z_I-z_c) & 0&0& (a_x+ \epsilon a_y) y\\
0& (1+z_c-a_z)(a_z + z_c-z_I)&(a_x- \epsilon a_y) y&0\\
0&(a_x- \epsilon a_y) y&(1-z_c-a_z)(a_z + z_I-z_c)&0\\
(a_x+ \epsilon a_y) y & 0&0&(1-z_c+a_z)(a_z + z_c-z_I)
\end{array}
\right].
\label{ellistate}
\end{align}}
The parameter $y$ stands for  $y=\sqrt{a_z^2+(z_I-z_c)^2}$. 

That the non-generic case of $a_z=0$ does not obtain as limit of \eqref{ellistate} is obvious. What may not be obvious is the fact that this limiting case exhibits an interesting, and perhaps surprising, feature of its own [see Remark 23 after Eq.\,\eqref{azeq}].\,$\blacksquare$

\section{Comparison with the work of Ali, Rau, and Alber \label{rausection}}
In this Section we contrast our approach and results with those of the famous work of Ali, Rau, and Alber (ARA)\,\cite{ali}, whose principal claim comprises two parts\,:
\begin{enumerate}[C1\,:]
\item Among all von Neumann measurements, either the horizontal or the vertical projection always yields the optimal classical correlation and quantum discord.
\item  The values thus computed remain optimal even when general POVM's are considered. 
\end{enumerate}

As for the second claim, the main text of ARA simply declares ``The Appendix shows how we may generalize to POVM  to get final compact expressions that are simple extensions of the more limited von Neumann measurements, thereby {\em yielding the same value} for the maximum classical correlation and discord.'' The Appendix itself  seems not to try enough towards validating this claim. It begins with ``Instead of von Neumann projectors, consider more general POVM. For instance, choose three orthogonal unit vectors mutually at $120^{\circ}$,
\begin{align}
\hat{s}_{0,1,2} = [\hat{z},(-\hat{z} \pm \sqrt{3} \hat{x})/2],
\label{raueq}
\end{align} 
and corresponding projectors $\cdots$.'' [It is not immediately clear how `orthogonal' is to be reconciled with `mutually at $120^{\circ}$' ]. Their subsequent reference to their Eq.\,(11) possibly indicates that ARA have in mind two more sets of such {\em three orthogonal unit vectors mutually at $120^{\circ}$} related to\,\eqref{raueq} through $SU(2)$ rotations. In the absence of concrete computation aimed at validating the claim, one is left to wonder if the second claim (C2) of ARA is more of an  assertion than deduction. 

We now know from our analysis in Section VI, however,  that the actual situation in respect of the second claim is much more subtle\,: the optimal three-element POVM is hardly of the {\em three orthogonal unit vectors mutually at $120^{\circ}$} type and, further, when a three-element POVM is required as the optimal one, there seems to be no basis to anticipate that it would yield `the same value for the maximum classical correlation and discord' as the one obtained from von Neumann. 

Admittedly, the present work is not the first to discover that ARA is not the last word on quantum discord of $X$-states. Several authors have pointed to examples of $X$-states which fail ARA\,\cite{zambrini11,zambriniepl,wang11,du12,chen11,adesso11,huang13}. But these authors have largely been concerned with the second claim (C2) of ARA. In contradistinction, our considerations below focuses exclusively on the first one (C1). 
In order that it be clearly understood as to what the ARA claim (C1) is not,  we begin with the following three statements\,:
\begin{enumerate}[S1\,:]
\item {\em If von Neumann projection proves to be the optimal POVM}, then the projection is either along the x or z direction.
\item von Neumann projection along the x or z direction {\em always proves to be the optimal POVM}.
\item von Neumann projection along either the x or z direction proves to be {\em the best among all von Neumann projections}. 
\end{enumerate}
\noindent
Our analysis has confirmed  that the first statement (S1) is absolutely correct. We also know that the second statement (S2) is correct except for a very tiny fraction of states corresponding to the wedge-like region between curves (1) and (2) in Fig.\,\ref{wedge}. 

The first claim (C1) of ARA corresponds, however, to neither of these two but to the third statement (S3). 
We begin with a counter-example to prove that this claim (S3) is non-maintainable. The example corresponds to
Mueller matrix 
\begin{align}
M = \left( \begin{array}{cccc}
1&0&0&0.23\\
0&0.76&0&0\\
0&0&0.6&0\\
0.3&0&0&0.8
\end{array}\right),
\label{ex1}
\end{align}
whose ellipsoid parameters are $(a_x,a_y,a_z,z_c) = (0.780936,0.616528,0.77183,0.122479)$. These parameters, together with $z_I=0.3$, fully specify the state in the canonical form.

\begin{figure}
\begin{center}
\scalebox{0.8}{\includegraphics{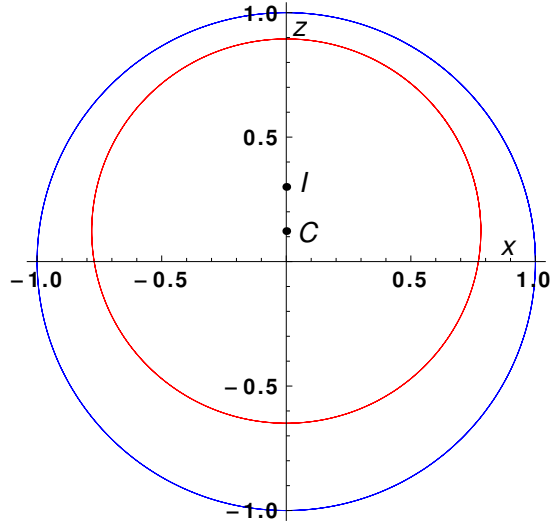}}
\end{center}
\caption{Showing the x-z section of the correlation ellipsoid associated with our example in Eq.\,\eqref{ex1}, with $z_I=0.3$, $z_c=0.122479$, and $(a_x,a_y,a_z) = (0.780936,0.616528,0.77183)$.
 \label{figx3}} 
\end{figure}

The parameter values verify the positivity requirements \eqref{cp1} and \eqref{cp2}. Further, it is seen that $M \in \Omega^+$ and corresponds to a nonseparable state. The x-z section of the correlation ellipsoid corresponding to this example is depicted in Fig.\,\ref{figx3}.

\begin{figure}
\begin{center}
\scalebox{0.8}{\includegraphics{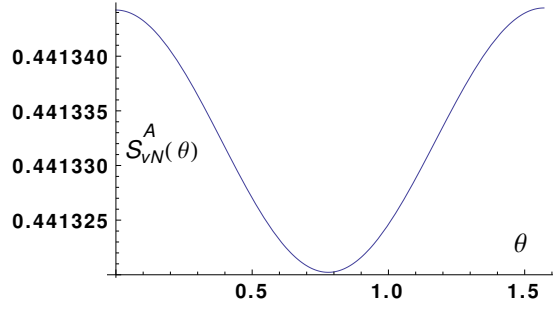}}
\end{center}
\caption{Showing the (average) conditional entropy $S^A_{vN}(\theta)$ resulting from von Neumann measurement $\Pi^{vN}_{\theta}$ for the case of our example in Eq.\,\eqref{ex1}. This example is so manufactured as to return the same value of $S^A_{vN}(\theta)$ for both x and z von Neumann and, further, to return a better $S^A_{vN}(\theta)$ for every other von Neumann.
 \label{figx4}} 
\end{figure}

Let us denote by $S^A_{vN}(\theta)$ the average conditional entropy post von Neumann measurement $\Pi^{vN}_{\theta}$ parametrized by angle $\theta$\,:
\begin{align}
\Pi^{vN}_{\theta} = \left\{ (1,\sin{\theta}, 0 , \cos{\theta})^T, \, (1,-\sin{\theta}, 0 ,-\cos{\theta})^T \right\}, ~~ 0 \leq \theta \leq \pi/2.
\end{align} 
It is clear that the output states are at distances $r(\theta),\,r^{\,'}(\theta)$ with respective conditional probabilities $p(\theta),\,p^{\,'}(\theta)$\,: 
\begin{align}
r(\theta) &= \frac{\sqrt{(m_{11} \,\sin{\theta})^2+(m_{30} + m_{33} \,\cos{\theta})^2}}{1+m_{03}\,\cos{\theta}},\nonumber\\
r^{\,'}(\theta) &= \frac{\sqrt{(m_{11} \,\sin{\theta})^2+(m_{30} - m_{33} \,\cos{\theta})^2}}{1-m_{03}\,\cos{\theta}};\nonumber\\
p(\theta) &= \frac{1+m_{03}\,\cos{\theta}}{2}, ~~ p^{\,'}(\theta) = \frac{1-m_{03}\,\cos{\theta}}{2}.
\end{align}
Thus $S^A_{vN}(\theta)$ evaluates to 
\begin{align}
S^A_{vN}(\theta) = \frac{1}{2} &\left[ S_2(r(\theta))+S_2(r^{\,'}(\theta)) \right. \nonumber\\
&~~~+ \left. m_{03}\,\cos{\theta}\,(S_2(r(\theta))-S_2(r^{\,'}(\theta)) )\right].
\label{von}
\end{align}
The behaviour of $S^A_{vN}(\theta)$ as a function of $\theta$ is shown in Fig.\,\ref{figx4}, and it is manifest that the optimal von Neumann obtains {\em neither at $\theta=0$ nor at $\pi/2$, but at $\theta=0.7792$ radians}. More strikingly, it is not only that neither $\theta=0$ or $\pi/2$  is the best, but both are indeed {\em the worst} in the sense that {\em von Neumann projection along any other direction returns a better value!} Thus this example renders the ARA claim (S3) untenable. 

\begin{figure}
\begin{center}
\scalebox{0.8}{\includegraphics{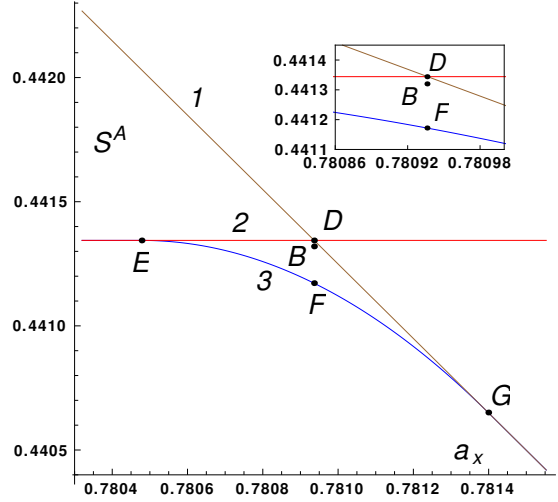}}
\end{center}
\caption{Showing variation of the mean conditional entropy $S^A$ as $a_x$ varies over the interval $[0.78032,0.781553]$, with $(a_y,a_z,z_c)$ fixed at $(0.616528,0.77183,0.122479)$ and $z_I$ at $0.3$. Curve (1) depicts $S^A_{vN}(\theta)$ for horizontal von Neumann measurement along $\theta=\pi/2$, the horizontal line [curve (2)] for vertical  von Neumann measurement along $\theta=0$, and curve (3) depicts $S^A_{\rm min}$ resulting from the three element (optimal) POVM $\Pi^{(3)}_{\theta}$. The example in Eq.\,\eqref{ex1} corresponds to $a_x=0.780936$. The inset compares the various measurement schemes for this example. D refers to the measurement being restricted to the ARA theorem of von Neumann projection along x and z-axis, B to the best von Neumann projection, and F to the optimal (three-element) measurement. 
 \label{figx5}} 
\end{figure}

Going beyond ARA, we  know from our analysis in Section\,\ref{saminsec} that if the von Neumann measurement indeed happens to be the optimal POVM, it can not obtain for any angle other than $\theta=0$ or $\pi/2$. Thus, the fact that the best von Neumann for the present example corresponds to neither angle is already a sure signature that a three-element POVM is lurking around as the optimal one for the state under consideration. Prompted by this signature, we embed the state under consideration in a one-parameter family with fixed $(a_z,a_y,z_c)=(0.616528,0.77183,0.122479)$ and $z_I=0.3$ but $a_x$ varying over the range $[0.7803,0.7816]$. The results are shown in Fig.\,\ref{figx5}.  
Curve (1) and curve (2) correspond respectively to the horizontal and vertical von Neumann projections, whereas curve (3) corresponds to the optimal three-element POVM. We emphasise that curve (3) is not asymptotic to curves (1) or (2), but joins them at $G$ and $E$ respectively. Our example of Eq.\,\eqref{ex1} embedded in this one-parameter family is highlighted by points $D,\, B,\,F$. This example is so manufactured that $S^A_{vN}(\theta)$ computed by the horizontal projection equals the value computed by the vertical projection, and is denoted by point $D$. The point $B$ corresponds to $S^A_{vN}(\theta)$ evaluated using the best von Neumann, which obtains for $\theta=0.779283$ radians as already noted, and $F$ to the value computed by the (three-element) optimal POVM. It may be noted that $D$ and $B$ are numerically quite close as highlighted by the inset. A numerical comparison of these values is conveniently presented in Table\,\ref{tablex1}. We see that D and B  differ only in the fifth place; even D and F differ only in the fourth place!
\begin{table}
\centering
\begin{tabular}{|c|c|c|c|}
\hline
Scheme&Elements& Optimal value & $S^A$\\
\hline
ARA & $\sigma_x$  or  $\sigma_z$ & equal & $0.441344$ \\
\hline
von Neumann & $\Pi^{vN}_{\theta}$, ~~$\theta \in [0,\pi/2]$ & $\theta_{\rm opt} = 0.779283$ & $0.44132$ \\
\hline
~~3-element POVM ~~&~~$\Pi^{(3)}_{\theta}$,~~ $\theta \in [0,\pi/2]$~~ & ~~$\theta_{\rm opt} =1.02158$ ~~&~~ $0.441172$~~\\
\hline
\end{tabular}
\caption{A comparison of the outcome of the ARA prescription with both the best von Neumann and the optimal three-element POVM for the example in Eq.\,\eqref{ex1}.\label{tablex1}}
\end{table}

\begin{figure}
\begin{center}
\scalebox{0.8}{\includegraphics{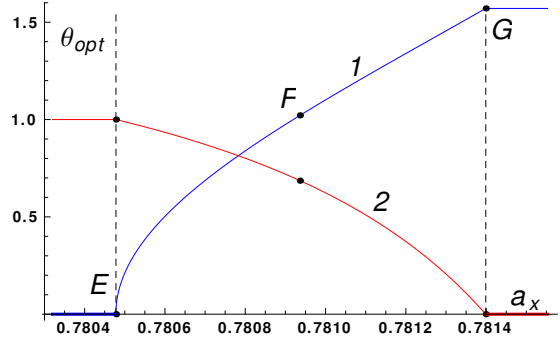}}
\end{center}
\caption{Showing the variation of optimal $\theta=\theta_{\rm opt}$ of $\Pi^{(3)}_{\theta}$ [curve (1)] leading to $S^A_{\rm min}$ depicted as curve (3) in Fig.\,\ref{figx5} as a function of $a_x$. Curve (2) shows the continuous evolution of the probability $p_0(\theta_{\rm opt})$ of the conditional state corresponding to input POVM element $(1,0,0,1)^T$, scaled by a factor of 2 to $2p_{0}(\theta_{\rm opt})$ for convenience. 
 \label{figx6}} 
\end{figure}

It is seen from Fig.\,\ref{figx5} that vertical von Neumann is the optimal POVM upto the point E (i.e. for $a_x \leq 0.780478$), from E all the way to G the three-element POVM $\Pi^{(3)}_{\theta}$  is the optimal one, and beyond G ($a_x \geq 0.781399$) the horizontal von Neumann is the optimal POVM. The {\em continuous evolution} of the parameter $\theta$ in $\Pi^{(3)}_{\theta}$ of Eq.\,\eqref{scheme} as one moves from E to G is shown in Fig.\,\ref{figx6}. Shown also is the {\em continuous} manner in which the probability $p_0(\theta)$ in Eq.\,\eqref{scheme} continuously varies from $0.5$ to zero as $a_x$ varies over the range from E to G.

In order to reconcile the manifest contradiction between ARA's first claim (S3 above) and our counter-example, we briefly reexamine their very analysis leading to the claim. As we shall see the decisive stage of their argument is  symmetry-based or group theoretical in tenor. It is therefore unusual that they carry around an extra baggage of irrelevant phase parameters, not only in the main text but also in the reformulation presented in their Appendix\,: the traditional first step in symmetry-based approach is to transform the problem to its simplest form (often called the canonical form) without loss of generality. Their analysis begins with parametrization of von Neumann measurements as [their Eq.\,(11)]
\begin{align}
B_i = V \Pi_i V^{\dagger}, ~~ i=0,1,
\end{align} 
where $\Pi_i = \proj{i}$ is the projector on the computation state $|i\rangle \in \{|0\rangle, |1\rangle \}$ and $V \in SU(2)$. With the representation $V = t 1\!\!1 + i \vec{y}.\vec{\sigma}$, $t^2+ y_1^2+y_2^2+y_3^2=1$ for $V \in SU(2)$ they claim that {\em three of these four} parameters $t,y_1,y_2,y_3$ are {\em independent}. Inspired by their Ref.\,[15] (our Ref.\,\cite{luo}), ARA recast  $t,y_1,y_2,y_3$ into four new parameters $m,n,k,\ell$ and once again emphasize that $k,m,n$ are three {\em independent} parameters describing the manifold of von Neumann measurements. \\

\noindent
{\bf Remark 17}\,: It is obvious that every von Neumann measurement on a qubit is fully specified by a pure state (the orthogonal state being automatically fixed), and hence the manifold of von Neumann measurements can be no larger than ${\cal S}^2$, the Bloch sphere. Indeed, this manifold is even `smaller'\,: it coincides with the real projective space ${\cal RP}^2 = {\cal S}^2/{\cal Z}_2$ of diameters in ${\cal S}^2$, since a pure state and its orthogonal partner  define {\em one and the same} von Neumann measurement. In any case, it is not  immediately clear in what sense could this two-manifold  be described by `three independent' parameters.\,$\blacksquare$\\

\noindent
{\bf Remark 18}\,: We should hasten to add, for completeness, that ARA did introduce subsequently, in an unusually well cited {\em erratum}\,\cite{alierratum}, another identity 
\begin{align}
m^2 + n^2 = klm
\label{araeq}
\end{align}
 which they claimed to be independent of  $t^2+ y_1^2+y_2^2+y_3^2=1$, and hence expected it to reduce the number of independent variables parametrizing the manifold of von Neumann measurements from three to two. To understand the structure of this new identity, which turns out to be of eighth degree in the original variables, define two complex numbers $\alpha = t-iy_3$, $\beta=y_1+iy_2$. Then ARA's definition of $k,\ell,m,n$ corresponds to  $k = |\alpha|^2$, $\ell = |\beta|^2$, $m = ({\rm Re} \alpha \beta)^2$, and $n= ({\rm Re} \alpha \beta)\, ({\rm Im} \alpha \beta)$ so that the ARA identity Eq.\,\eqref{araeq} reads 
\begin{align}
({\rm Re} \alpha \beta)^4 + ({\rm Re} \alpha \beta)^2 \,({\rm Im} \alpha \beta) = |\alpha\beta|^2 ({\rm Re} \alpha \beta)^2,
\label{araeqb}
\end{align}
showing that it is indeed independent of $t^2 + y_1^2+y_2^2+y_3^2=k+ \ell=1$ as claimed by ARA. Indeed, it is simply the Pythagorean theorem $|z|^2 = ({\rm Re} z)^2 +({\rm Im} z)^2 $ valid for any complex number $z$  puffed up to the appearance of a sophisticated eighth degree real homogeneous form. It is unlikely that such an {\em universal} identity, valid for any four real numbers, would ever aid in reducing the number of independent parameters. Not only ARA, but also the large number of works which cite this erratum, seem to have missed this universality aspect of the ARA identity\,\eqref{araeq}.\,$\blacksquare$

Returning now to the clinching part of the ARA analysis, after setting up the expression for the conditional entropy as a function of their independent variables $k,m,n$ they correctly note that it could be minimized ``by setting equal to zero its partial derivatives with respect to $k,m$ and $n$.'' Rather than carrying out this step, however, they prefer a short cut in the form of a symmetry argument. They `{\em observe}' that the problem has a symmetry (this is the symmetry of inversion about the z-axis which we used in Section\,\ref{optsec} to simplify our optimization problem), and then use the {\em unusual symmetry argument} that if a problem has a symmetry its solution {\em ought to be} invariant under that symmetry. Obviously, one knew in advance that the only von Neumann projections that are invariant under the symmetry under consideration are the (vertical or) z-projection and the horizontal projection, the latter meaning x or y-projection according as $a_x > a_y$ or $a_y > a_x$. This version of symmetry argument is unusual, since the familiar folklore version reads\,: {\em if a problem has a symmetry, its solution ought to be covariant (and not necessarily invariant) under the symmetry}. In any case, unless  the ARA version of symmetry argument 
be justified as arising from some {\em special aspect} of the problem under consideration,  its deployment would amount to assuming {\em a priori} that either z or x-projection is the best von Neumann; but then such an assumption would  precisely 
amount to the claim (S3) ARA set out to prove as the very central theorem of their work. {\em The point being made is that all generic $X$-states (i.e., states for which $a_y \neq a_x$) share the same ${\cal Z}_2 \times {\cal Z}_2$ symmetry; and therefore it would seem that the ARA argument anchored on this symmetry either stands for all $X$-states or for none}.\\

\noindent
{\bf Remark 19}\,: The ARA version of symmetry argument would remain justified if it were the case that the problem is expected, from other considerations, to have a {\em unique} solution. This could happen, for instance, in the case of  {\em convex optimization}. But von Neumann measurements {\em do not form a convex set}, and hence the ARA problem of optimization over von Neumann measurement is not one of convex optimization. Thus demanding a unique solution in their case would again amount to an {\em a priori} assumption equivalent to the theorem they set out to prove. It is for this subtle aspect which probably lies at the root of ARA failure.\,$\blacksquare$

\section{$X$-states with vanishing discord}
Some authors have earlier considered methods to enumerate the zero discord $X$-states\,\cite{zero-disc,chrus10,chrus12}. 
Our analysis below is directly based on the very definition of vanishing discord and hence is elementary in nature; more importantly, it leads to an {\em exhaustive} classification of these states, correcting an earlier claim\,\cite{chrus12}. Any  two-qubit state of vanishing quantum discord can be written as\,\cite{modi-rmp} 
\begin{align}
\hat{\rho}_{AB} = U_A |0\rangle \langle 0| U_A^{\dagger} \otimes p_1 \hat{\rho}_{B1} + U_A |1\rangle \langle 1| U_A^{\dagger} \otimes p_2 \hat{\rho}_{B2}, 
\label{zerod}
\end{align}
with $p_1,\,p_2 \geq0$, $p_1 + p_2 =1$, {\em the measurements being now assumed performed on subsystem A}. We may write
\begin{align}
p_1 \hat{\rho}_{B1} &= \begin{bmatrix}
a_1 & b_1 \\
b_1^* & c_1
\end{bmatrix}, ~~
p_2 \hat{\rho}_{B2} = \begin{bmatrix}
a_2 & b_2 \\
b_2^* & c_2
\end{bmatrix}, \nonumber\\~~ 
U_A &= \begin{bmatrix}
\alpha & \beta\\
-\beta^* & \alpha^*
\end{bmatrix} \in\,SU(2), ~ |\alpha|^2 + |\beta|^2=1.
\end{align}
Clearly, the reduced state of subsystem B is $p_1 \hat{\rho}_{B1} + p_2 \hat{\rho}_{B2}$, and that of $A$ equals $ U_A\, (p_1 |0\rangle \langle 0| + p_2 |1 \rangle \langle 1|)\,U_A^{\dagger}$. We should next reconcile this nullity condition with the demand that the state under consideration be an $X$-state in the canonical form\,\eqref{xcan}.  From the off-diagonal blocks of $\hat{\rho}_{AB}$ we immediately see that $a_1 = a_2$ and $c_1 =c_2$. Then
${\rm Tr}\,\hat{\rho}_{AB} = a_1 + c_1 + a_2 + c_2 = 1$ implies $a_1 + c_1 = 1/2= a_2+c_2$. 
Vanishing of the $01$ and $23$ elements of $\hat{\rho}_{AB}$ forces the following constraints\,:
\begin{align}
&|\alpha|^2 b_1 + |\beta|^2 b_2 = 0,\nonumber \\
&|\alpha|^2 b_2 + |\beta|^2 b_1 = 0.
\end{align}     
These imply in turn that either $|\alpha| = |\beta| = 1/\sqrt{2}$ or $b_1 = b_2=0$. The first case of $|\alpha| = |\beta| = 1/\sqrt{2}$ forces $b_2 = -b_1$, and we end up with a two-parameter family of zero discord states in the canonical form\,:
\begin{align}
\hat{\rho}^A(a,b) &= \frac{1}{4}\left( 
\begin{array}{cccc}
1+a& 0 &0 & b\\
0& 1 -a & b & 0\\
0&b&1+a &0\\
b&0&0&1-a
\end{array}
\right)\nonumber\\
&= \frac{1}{4} \left[\sigma_0 \otimes \sigma_0 + a \sigma_0 \otimes \sigma_3 + b \sigma_1 \otimes \sigma_1\right]. 
\label{onewaya}
\end{align}
Positivity of $\hat{\rho}^A(a,b)$ places the constraint $a^2 + b^2 \leq 1$, a disc in the $(\sigma_0 \otimes \sigma_3, \sigma_1 \otimes \sigma_1)$ plane.  The special case $b=0$ corresponds to the product state 
\begin{align}
\hat{\rho}^{AB}(a) = \frac{1}{4} 1\!\!1 \otimes \begin{bmatrix}  1+a & 0 \\ 0& 1-a \end{bmatrix}.
\end{align}

If instead the measurement was performed on the $B$ subsystem, then it can be easily seen that  similar arguments can be used to arrive at the following form for zero discord states\,: 
\begin{align}
\hat{\rho}^B(a,b) &= \frac{1}{4}\left( 
\begin{array}{cccc}
1+a & 0 &0 & b\\
0& 1+a & b & 0\\
0&b&1-a&0\\
b&0&0&1-a
\end{array}
\right),\nonumber\\
& = \frac{1}{4}\left[ \sigma_0 \otimes \sigma_0 + a \sigma_3 \otimes \sigma_0 + b \sigma_1 \otimes \sigma_1\right]. 
\label{onewayb}
\end{align}
Positivity again constrains $a,b$ to the disc $a^2 + b^2 \leq 1$, this time in the $(\sigma_3 \otimes \sigma_0, \sigma_1 \otimes \sigma_1)$ plane. 

The intersection between these two two-parameter families comprises the one-parameter family of $X$-states
\begin{align}
\hat{\rho}_{AB} = \frac{1}{4}\left[ \sigma_0 \otimes \sigma_0 + b \sigma_1 \otimes \sigma_1\right], ~~ -1 \leq b \leq 1. 
\label{onewayint}
\end{align}
{\em But these are  not the only two-way zero discord states}, and this fact is significant in the light of the claims of\,\cite{chrus12}.  To see this, note that in deriving the canonical form \eqref{onewaya} we assumed $\beta \neq 0$. So we now consider the case $\beta=0$, so that \eqref{zerod} reads
\begin{align}
\hat{\rho}_{AB} = p_1 |0\rangle \langle 0| \otimes \hat{\rho}_{B1} + p_2 |1\rangle \langle 1| \otimes \hat{\rho}_{B2}.
\end{align}
The demand that this be an $X$-state forces $\hat{\rho}_{AB}$ to be diagonal in the computational basis\,:
\begin{align}
\hat{\rho}_{AB}(\{p_{k \ell} \}) = \sum_{k,\ell=0}^1 p_{k \ell}\, |k\rangle \langle k| \otimes |\ell \rangle \langle \ell|.
\label{twoway}
\end{align}
It is manifest that all $X$-states of this three-parameter family, determined by probabilities $\{p_{k\ell} \}$, $\sum p_{k\ell} =1$, and worth a {\em tetrahedron} in extent, have vanishing quantum discord and, indeed, vanishing two-way quantum discord.  

The intersection of \eqref{onewaya} and \eqref{onewayb} given in  \eqref{onewayint} is not `really' outside the tetrahedron \eqref{twoway} in the canonical form because it can be diagonalized by a local unitary $U_A \otimes U_B$, $U_A= U_B = \exp [-i \pi \sigma_2/4]$\,:
\begin{align}
\sigma_0 \otimes \sigma_0 + b \sigma_1 \otimes \sigma_1 \to  \sigma_0 \otimes \sigma_0 + b \sigma_3 \otimes \sigma_3. 
\end{align}
Stated differently, the family of  one-way (but not two-way) zero discord $X$-states in the canonical form is not a disc, but a disc with the diameter removed. \\

\noindent
{\bf Remark 20}\,: Strictly speaking, this is just an half disc with diameter removed, as seen from the fact that in \eqref{onewaya}, \eqref{onewayb}, and \eqref{onewayint} the two states $(a,b)$, $(a,-b)$ are local unitarily equivalent under $U_A \otimes U_B = \sigma_0 \otimes \sigma_3$ or $\sigma_3 \otimes \sigma_0$.\,$\blacksquare$

We now consider the nullity signature of these states on the associated correlation ellipsoids. For the one-way zero discord states in \eqref{onewayb} the non-zero Mueller matrix entries are  
\begin{align}
m_{30} = a, ~~ m_{11} = b. 
\end{align}
This ellipsoid is actually a symmetric line segment parallel to the  x-axis, of extent $2b$, translated by extent $a$, perpendicular to the line segment (i.e., along z)\,: $\{(x,y,z) = (x,0,a)|-b\leq x \leq b \}$;  it is symmetric under reflection about the z-axis. Equivalently, {\em the bisector of this line segment is radial}. For measurements on the A side we have from \eqref{onewaya} 
\begin{align}
m_{03} = a, ~~ m_{11} = b,
\end{align}
and we get the same line segment structure (recall that now we have to consider $M^T$ in place of $M$). 

For the two-way zero discord states\,\eqref{twoway} we have  
\begin{align}
m_{03} & = p_{00}-p_{01}+p_{10} - p_{11},\nonumber\\
m_{30} &= p_{00}+p_{01}-p_{10}- p_{11},\nonumber\\
m_{33} &= p_{00}-p_{01}-p_{10} + p_{11},
\end{align}
corresponding to a point in the tetrahedron. We note that the associated correlation ellipsoid is a line segment of a diameter {\em shifted along the diameter itself}. That is, the line segment itself is radial. While the extent of the line segment and the shift are two parameters, the third parameter is the image ${\rm I}$ of the maximally mixed input, which does not contribute to the `shape' of the ellipsoid, but does contribute to the state. This three parameter family should be contrasted with the claim of\,\cite{chrus12} that an `$X$-state is purely classical if and only if $\hat{\rho}_{AB}$ has components only along $\sigma_0 \otimes \sigma_0, \sigma_1 \otimes \sigma_1 $', thereby implying a one-parameter family.

\section{States not requiring an optimization}
We now exhibit a large class of states for which one can write down {\em analytic expression} for quantum discord simply by inspection, without the necessity to perform explicit optimization over all measurements. This class is much larger than the one studied by Luo\,\cite{luo}. \\

\noindent
{\bf Centered states ($z_c=0$)}\,:  
Consider $X$-states for which the associated correlation ellipsoid is {\em centered at the origin}: 
\begin{align}
z_c = &~~\frac{m_{30} - m_{03}m_{33}}{1-m_{03}^2} =0, \nonumber\\
{\rm i.e.},~~  &~m_{30} = m_{03}m_{33}.
\label{centered}
\end{align}
This implies on the one hand that only two of the three Mueller matrix elements $m_{03},\,m_{30},\,m_{33} $ are independent. On the other hand, it implies that the product of $m_{03}, m_{30}, m_{33}$ is necessarily positive and thus, by local unitary, all the three can be arranged to be positive without loss of generality. Let us take $m_{03} = \sin{\theta} > 0$, then we have $m_{30} = m_{33}\, \sin{\theta}$. 
So, we now have in the canonical form a four-parameter family of Mueller matrices
\begin{align}
M(\gamma_1,\,\gamma_2,\,\gamma_3;\,\theta) = \begin{bmatrix}
1 & 0 & 0 & \sin{\theta} \\
0& \gamma_1 \cos{\theta} & 0 & 0 \\
0& 0& \gamma_2 \,\cos{\theta} & 0 \\
\gamma_3 \, \sin{\theta} & 0 & 0 & \gamma_3
\end{bmatrix},
\label{center}
\end{align}
and correspondingly a three-parameter family of correlation ellipsoids centered at the origin, with principal axes $(a_x,a_y,a_z) = (\gamma_1, |\gamma_2|, \gamma_3)$, $z_c=0$, and $z_I = \gamma_3 \, \sin{\theta}$. We continue to assume $|\gamma_2| \leq \gamma_1$.\\
 
\noindent
{\bf Remark 21}\,: Note that we are not considering here just the case of Bell-diagonal states, which too correspond to ellipsoids centered at the origin. In the Bell-diagonal case, the point {\rm I}, which represents $\hr{A}$, is located at the origin and, as an immediate consequence, $S^A_{\rm min}$ is obviously determined by the major axis of the ellipsoid. In the present case, $z_I = \gamma_3 \,\sin{\theta} \neq 0$, and $S^A_{\rm min}$ does depend on $z_I$. Indeed, the case of Bell-diagonal states corresponds to $\sin{\theta}=0$, and hence what we have here is a one-parameter generalization of the Bell-mixture case.\,$\blacksquare$

The  four parameter family of density matrices corresponding to Eq.\,\eqref{center} takes the form 
\begin{align}
&\rho(\gamma_1,\,\gamma_2,\,\gamma_3;\,\theta)=  \nonumber\\
&  \frac{1}{4}
\begin{bmatrix}
(1+\gamma_3)(1+\sin{\theta}) & 0 & 0 & (\gamma_1 + \gamma_2)\,\cos{\theta}\\
0 & (1-\gamma_3)(1-\sin{\theta}) & (\gamma_1 - \gamma_2)\,\cos{\theta}& 0 \\
0&(\gamma_1 - \gamma_2)\,\cos{\theta} & (1-\gamma_3)(1+\sin{\theta})&0\\
(\gamma_1 + \gamma_2)\,\cos{\theta} & 0 & 0 & (1+\gamma_3)(1-\sin{\theta})
\end{bmatrix}.
\label{circ}
\end{align}
The first CP  condition \eqref{cp1} reads
\begin{align}
 ~~(1+\gamma_3)^2 &- \sin^2{\theta}  (1+\gamma_3)^2 \geq (\gamma_1 + \gamma_2)^2 \,\cos^2{\theta},\nonumber\\
{\rm i.e.},  &~~  \gamma_1 + \gamma_2 - \gamma_3 \leq 1,
\end{align} 
while the second CP condition \eqref{cp2} reads
\begin{align}
~~(1-\gamma_3)^2 &- \sin^2{\theta}  (1-\gamma_3)^2 \geq (\gamma_1 - \gamma_2)^2 \, \cos^2{\theta}\nonumber\\
{\rm i.e.}, &~~ \gamma_1 + \gamma_3 -  \gamma_2 \leq 1.  
\end{align}
Recalling that $|\gamma_2| \leq \gamma_1$, these two conditions can be combined into a {\em single CP condition} 
\begin{align}
\gamma_1 + |\gamma_3 -  \gamma_2|\leq 1.
\end{align}\\

\noindent
{\bf Remark 22}\,: We note a special property of these states in respect of quantum discord.
As noted in the Remark 12 after Eq.\,\eqref{threerange}, {\em for centered correlation ellipsoids the optimal POVM is von Neumann projection along either $x$ or $z$ according as $a_x > a_z$ or $a_x<a_z$}. There is no need for `optimization' in this case of centered ellipsoids.\,$\blacksquare$  

Having thus fully characterized $X$-states corresponding to centered correlation ellipsoids, 
and having exposed their triviality in respect of `optimization', we may now look at some particular cases. \\

\noindent
{\bf Circular states ($a_x=a_z>a_y$, $z_c=0$)}\,: This special case of centered states corresponds to  $\gamma_3 = \gamma_1$.
For this class of states, every von Neumann measurement in the x-z plane (indeed, any POVM with all the measurement elements lying in the x-z plane) is equally optimal. The fact that location of ${\rm I}$  plays no role in determining the optimal POVM or $S^A_{\rm min}$ for this class of $X$-states is pictorially obvious, independent of our analysis in Section VI.

The four eigenvalues of $\hat{\rho}(\gamma_1,\,\gamma_2;\,\theta)$ of \eqref{circ} are
\begin{align}
\{\lambda_j \} &= \frac{1}{4} \left\{ 1 + \epsilon \gamma_1 \pm y \right\},\nonumber\\
y &=  \sqrt{(1 + \epsilon \gamma_1)^2 \sin^2{\theta} + (\gamma_1 + \epsilon \gamma_2 )^2 \cos^2{\theta}} ,
\label{eigencirc}
\end{align} 
$\epsilon$ being a signature. We can explicitly write down the various quantities of interest in respect of the circular states $\hat{\rho} (\gamma_1,\gamma_2;\theta)$.  First, we note that the conditional entropy post measurement is simply the entropy of any  state  on the boundary circle in the x-z plane, and hence 
\begin{align}
S^A_{\rm min}= S_2(m_{33})=S_2(\gamma_1).
\end{align}
By Eqs.\,\eqref{ents} and \eqref{mi}, we have 
\begin{align}
I(\hat{\rho}(\gamma_1,\,\gamma_2;\,\theta))&= S_2(\gamma_1\,\sin{\theta}) + S_2(\sin{\theta}) - S(\{\lambda_j\}),\nonumber\\
C(\hat{\rho}(\gamma_1,\,\gamma_2;\,\theta)) &= S_2(\gamma_1\,\sin{\theta}) - S_2(\gamma_1),\nonumber\\
D(\hat{\rho}(\gamma_1,\,\gamma_2;\,\theta))&= S_2(\sin{\theta}) + S_2(\gamma_1) - S(\{\lambda_j\}),
\label{corrcirc}
\end{align}
where $\lambda_j \equiv \lambda_j(\gamma_1,\,\gamma_2;\,\theta)$ are given in Eq.\,\eqref{eigencirc}. 
Finally, we note that with the local unitary freedom, this 3-parameter class of states can be lifted to a 9-parameter family of states.\\

\noindent
{\bf Spherical states ($a_x=a_z=a_y$, $z_c=0$)}: The correlation ellipsoid corresponding to these states is a sphere with $z_c=0$. They can be obtained as a subset of circular states by setting $\gamma_1 = |\gamma_2|$. The expressions for the correlation are the same as those of circular states as given in Eq.\,\eqref{corrcirc}. We note that, the spherical states form a 2-parameter family of states in the canonical form. We can lift this family, using local unitaries, to a seven parameter family of states, five parameters coming from the local unitary transformations, one local  parameter having been `lost' owing to the degeneracy $m_{11} = |m_{22}|$ for spherical states. \\

\noindent
{\bf Bell mixtures ($z_c=0=z_I$)}\,:
Another particular case of centered states is the convex mixture of Bell-states which was the centre of the study in\,\cite{luo}. It corresponds to the additional restriction $z_I=0$ on the general centered $X$-states. We can write the state as 
$\hat{\rho} = \sum_{j=1}^4 p_j |\phi_j \rangle \langle \phi_j|$, i.e.,
\begin{align}
\hat{\rho} = \frac{1}{2} \left( \begin{array}{cccc}
p_1 + p_2 & 0 & 0 & p_1 - p_2 \\
0& p_3+p_4 & p_3 -p_4 &0 \\
0& p_3 - p_4 & p_3 +p_4 & 0\\
p_1 - p_2 &0&0& p_1 + p_2
\end{array}
\right).
\end{align} 
The corresponding Mueller matrix is diagonal with 
\begin{align}
m_{11} &= p_1 + p_3 - (p_2 + p_4), \nonumber\\
m_{22} & =p_1 + p_4 -(p_2 + p_3), \nonumber\\
m_{33} &= p_1 + p_2 - (p_3 + p_4).
\end{align}
That the optimal measurement is a von Neumann projection along the direction of the longest principal axis of the ellipsoid is pictorially obvious in this case, and this fact was the focus of the detailed analysis of Luo\,\cite{luo}. \\

\noindent
{\bf Linear states ($a_z=0$)}\,:
Another example of states for which the quantum discord can be immediately written down is the class of states whose x-z section of the correlation ellipsoid is a line segment. We call these states  linear states, and they correspond to $a_z=0$. We do not assume $z_c=0$, and so this family is {\em not a subset of the family of centered states}. The condition $a_z=0$ demands 
\begin{align}
m_{33} = m_{03} m_{30}.
\label{azzero}
\end{align} 
As in the case of \eqref{centered} $m_{33},\,m_{03},\,m_{30}$ can all be arranged to be nonnegative without loss of generality. Since ${\rm det}\,M=0$, the ellipsoid is in the intersection $\Omega^+ \bigcap \Omega^-$. The correlation ellipsoid is an elliptic disc whose centre is radially offset by an extent $z_c$ from the centre of the Bloch ball, to a plane parallel to the x-y plane.  We thus have a  four parameter family of states in the canonical form, and the Mueller matrix for this family can be parametrized as
\begin{align}
M(\gamma_1,\gamma_2,\gamma_3,\theta) = \left[ 
\begin{array}{cccc}
1 & 0 & 0 & \sin{\theta}\\
0& \gamma_1 \cos{\theta} & 0 & 0\\
0 & 0& \gamma_2 \cos{\theta} &0\\
\gamma_3 & 0 & 0 & \gamma_3 \sin{\theta}
\end{array}
\right],
\end{align}
where we continue to assume $|\gamma_2| < \gamma_1$. The parameters of the correlation ellipsoid is given by $(a_x,a_y,a_z,z_c)=(\gamma_1,|\gamma_2|,0,\gamma_3)$, with $z_I=\gamma_3=z_c$. Then the CP conditions demand that $\gamma_1 + |\gamma_2| \leq \sqrt{1-\gamma_3^2}$. So we have $|\gamma_2| \leq {\rm min}\, (\gamma_1, \sqrt{1-\gamma_3^2} - \gamma_1)$,  and this condition subsumes $|\gamma_2| < \gamma_1$ but could prove to be stronger.

The  four parameter family of density matrices (in the canonical form) corresponding to Eq.\,\eqref{azzero} takes the form 
\begin{align}
&\rho(\gamma_1,\,\gamma_2,\,\gamma_3;\,\theta)=  \nonumber\\
&  \frac{1}{4}
\begin{bmatrix}
(1+\gamma_3)(1+\sin{\theta}) & 0 & 0 & (\gamma_1 + \gamma_2)\,\cos{\theta}\\
0 & (1+\gamma_3)(1-\sin{\theta}) & (\gamma_1 - \gamma_2)\,\cos{\theta}& 0 \\
0&(\gamma_1 - \gamma_2)\,\cos{\theta} & (1-\gamma_3)(1+\sin{\theta})&0\\
(\gamma_1 + \gamma_2)\,\cos{\theta} & 0 & 0 & (1-\gamma_3)(1-\sin{\theta})
\end{bmatrix}.
\end{align}
That
\begin{align}
S^A_{\rm min} = S_2(\sqrt{\gamma_1^2+\gamma_3^2})
\label{azeq}
\end{align}
is pictorially obvious from the fact that the correlation ellipsoid is an elliptic disc with major axis parallel to the x-axis. \\

\noindent
{\bf Remark 23}\,:
The role of the larger invariance group $SO(3,1)$ in this case has an unusual aspect to it. On the Mueller matrix M, the effect of $SO(3,1)$ action is seen to run the angle $\theta$ over the range $(-\pi/2,\pi/2)$. But the ellipsoid itself has no dependence on this $SO(3,1)$ action or on the angle $\theta$. Not even $z_I$ has any dependence on the $SO(3,1)$ element. Normally, $SO(3,1)$ action would have driven $z_I$ freely over the range $(z_c-a_z,z_c+a_z)$. Since $a_z=0$, the degree of freedom $z_I$ in the present case is frozen or compactified to the single point $z_c$. Thus, to reconstruct the state from the correlation ellipsoid, we need the $SO(3,1)$ element being supplied as an additional piece of information (over and above $z_I$).\,$\blacksquare$

\begin{widetext}
\subsection*{Appendix\,: Matrix elements of $\rho$ and $M$}
Matrix elements of $\rho_{AB}$ in terms of the Mueller matrix elements is given by\,: 
{\scriptsize
\begin{align*}
\rho_{AB} = \frac{1}{4}
\left[
\begin{array}{c|c|c|c}
m_{00} + m_{03} + m_{30} + m_{33} &m_{01} + i m_{02} + m_{31} + i m_{32} & m_{10} - i m_{20} + m_{13} - i m_{23} & m_{11}+i m_{12} - i m_{21}+m_{22} \\
\hline
m_{01} - i m_{02} + m_{31} - i m_{32}&m_{00} - m_{03} + m_{30} - m_{33}&m_{11}-i m_{12} - i m_{21}-m_{22}&m_{10} - i m_{20} - m_{13} + i m_{23}\\
\hline
m_{10} + i m_{20} + m_{13} + i m_{23}&m_{11}+i m_{12} + i m_{21}-m_{22}&m_{00} + m_{03} - m_{30} - m_{33}&m_{01} + i m_{02} - m_{31} - i m_{32}\\
\hline
m_{11}-i m_{12} + i m_{21}+m_{22}&m_{10} + i m_{20} - m_{13} - i m_{23} &m_{01} - i m_{02} - m_{31} + i m_{32}&m_{00} - m_{03} - m_{30} + m_{33}
\end{array} 
\right],
\end{align*}}
and that of $M$ in terms of ${\rho}_{AB}$ by\,:
{\scriptsize 
\begin{align*}
M = \left[
\begin{array}{c|c|c|c}
1& \rho_{01} + \rho_{10} + \rho_{23} + \rho_{32} & -i[(\rho_{01} - \rho_{10}) + (\rho_{23} + \rho_{32})] & (\rho_{00}  - \rho_{11}) + (\rho_{22} - \rho_{33})\\
\hline
\rho_{02} + \rho_{20} + \rho_{13} + \rho_{31} & \rho_{03} + \rho_{30} + \rho_{12} + \rho_{21} & -i [(\rho_{03} - \rho_{12}) + (\rho_{21} - \rho_{30})] & \rho_{02} + \rho_{20} - (\rho_{13} + \rho_{31}) \\ 
\hline
i [(\rho_{02} -\rho_{20})+ (\rho_{13} - \rho_{31})] & i [(\rho_{03} + \rho_{12}) - (\rho_{21} + \rho_{30})] & \rho_{30} + \rho_{03} - (\rho_{12} + \rho_{21}) & i[(\rho_{02} - \rho_{20}) - (\rho_{13} - \rho_{31})] \\
\hline
\rho_{00} + \rho_{11} -( \rho_{22} + \rho_{33}) & \rho_{01} + \rho_{10} -( \rho_{23} + \rho_{32}) & -i[(\rho_{01} -\rho_{10}) - (\rho_{23} - \rho_{32})] & \rho_{00} + \rho_{33} -( \rho_{11} + \rho_{22})
\end{array}
\right]. 
\label{app}
\end{align*}
}
\end{widetext}

\noindent
{\em Note}\,: This work was presented in the AQIS13 Conference, Chennai, India.


\begin{thebibliography}{}
\bibitem{horo-rmp} R. Horodecki, P. Horodecki, M. Horodecki, K. Horodecki, Rev. Mod. Phys. {\bf 81}, 865 (2009). 

\bibitem{modi-rmp} K. Modi, A. Brodutch, H. Cable, T. Paterek, V. Vedral,  Rev. Mod. Phys.  {\bf 84}, 1655 (2012).

\bibitem{celeri-rmp} L. C. Celeri, J. Maziero, R. M. Serra,  Int. J. Quantum Inform. {\bf 9}, 1837 (2011).

\bibitem{disc-imp1} 
A. Datta, A. Shaji, and C. M. Caves,  Phys. Rev. Lett. {\bf 100}, 050502 (2008);
B. P. Lanyon, M. Barbieri, M. P. Almeida, and A. G. White, \prl {\bf 101}, 200501 (2008); 
D. Cavalcanti, L. Aolita, S. Boixo, K. Modi, M. Piani, and A. Winter,  \pra {\bf 83}, 032324 (2011).
V. Madhok and A. Datta, Int. J. Mod. Phys. B {\bf 27}, 1345041 (2013). 

\bibitem{disc-imp2}
M. Piani, S. Gharibian, G. Adesso, J. Calsamiglia, P. Horodecki, and A. Winter,  \prl {\bf 106}, 220403 (2011);
M. F. Cornelio, M. C. de Oliveira, and F. F. Fanchini, \prl {\bf 107}, 020502 (2011).
M. Gu et al.,  Nat. Phys. {\bf 8}, 671 (2012); 
B. Dakic et al., Nat. Phys. {\bf 8}, 666 (2012).


\bibitem{disc-imp3}
K. K. Sabapathy, J. S. Ivan, S. Ghosh, and R. Simon, arXiv:1304.4857 [quant-ph].


\bibitem{xstates1} 
T. Yu and J. H. Eberly, Quantum Inf. Comput. {\bf 7}, 459 (2007);
A. R. P. Rau, J. Phys. A: Math. Theor. {\bf 42}, 412002 (2009);
M. D. Lang and C. M. Caves Phys. Rev. Lett. 105 150501 (2010);
N. Quesada, A. Al-Qasimi, and D. V. James, J. Mod. Opt. {\bf 59}, 1322 (2012).



\bibitem{xstates2}
B. Bellomo, R. Lo Franco, and G. Compagno, Phys. Rev. A {\bf 77}, 032342 (2008);
R. Dillenschneider, Phys. Rev. B {\bf 78}, 224413 (2008); 
T. Werlang, C. Trippe, G. A. P. Ribeiro, and G. Rigolin, Phys. Rev. Lett. {\bf 105}, 095702 (2010);
J. P. G. Pinto, G. Karpat, and F. F. Fanchini,  Phys. Rev. A {\bf 88}, 034304 (2013).


\bibitem{luo} S. Luo, \pra {\bf 77}, 042303 (2008).

\bibitem{ali} M. Ali, A. R. P. Rau, and G. Alber, \pra {\bf 81}, 042105 (2010). 


\bibitem{wang11} X. Lu, J. Ma, Z. Xi, and X. Wang, \pra {\bf 83}, 012327 (2011). 

\bibitem{chen11} Q. Chen, C. Zheng, S. Yu, X. X. Yi, and C. H. Oh, \pra {\bf 84}, 042313 (2011).



\bibitem{du-geom} M. Shi, F. Jiang, C. Sun, and J. Du, New J. Phys. {\bf 13}, 073106 (2011). 

\bibitem{du12} M. Shi, C. Sun, F. Jiang, X. Yan, and J. Du, \pra {\bf 85}, 064104 (2012).

\bibitem{huang13} Y. Huang, \pra {\bf 88}, 014302 (2013).


\bibitem{james11} A. Al-Qasimi and D. F. V. James, \pra {\bf 83} 032101 (2011). 

\bibitem{adesso11} D. Girolami and G. Adesso, \pra {\bf 83}, 052108 (2011).

\bibitem{zambrini11} F. Galve, G. L. Giorgi, and R. Zambrini, \pra {\bf 83}, 012102 (2011); F. Galve, G. L. Giorgi, and R. Zambrini, \pra {\bf 83}, 069905(E) (2011).

\bibitem{zambriniepl} F. Galve, G. L. Giorgi, and R. Zambrini, EPL {\bf 96}, 40005 (2011).

\bibitem{adesso11b} D. Girolami, M. Paternostro, and G. Adesso, J. Phys. A: Math. Theor. {\bf 44}, 352002 (2011).

\bibitem{nassajour13} S. J. Akhtarshenas, H. Mohammadi, F. S. Mousavi, and V. Nassajour, arXiv:1304.3914 (quant-ph). 




\bibitem{simon-mueller1}
R. Simon, Opt. Commun. {\bf 42}, 293 (1982).

\bibitem{simon-mueller2} 
M. S. Kumar and R. Simon, Opt. Commun. {\bf 88}, 464 (1992);
R. Sridhar and R. Simon, J. Mod. Opt. {\bf 41}, 1903 (1994).

\bibitem{simon-mueller3}
B. N. Simon, S. Simon, F. Gori, M. Santarsiero, R. Borghi, N. Mukunda, and R. Simon, Phys. Rev. Lett. {\bf 104}, 023901 (2010).

\bibitem{simon-mueller4}
B. N. Simon, S. Simon, N. Mukunda, F. Gori, M. Santarsiero, R. Borghi, and R. Simon, J. Opt. Soc. Am. A {\bf 27}, 188 (2010).  	




\bibitem{zurek01} H. Ollivier and W. H. Zurek,  \prl {\bf 88}, 017901 (2001).

\bibitem{vedral01} L. Henderson and V. Vedral, J. Phys. A: Math. Gen. {\bf 34}, 6899 (2001).

\bibitem{winter05} B. Groisman, S. Popescu, and A. Winter, \pra {\bf 72}, 032317 (2005). 

\bibitem{neillbook} See for instance, E. L. O'Neill, {\em Introduction to Statistical Optics}, Dover Publications (2004); E. Wolf, {\em Introduction to the Theory of Coherence and Polarization of Light}, Cambridge University Press (2007).  

\bibitem{cj} A. Jamio\l kowski, Rep. Math. Phys. {\bf 3}, 275 (1972); M.-D. Choi, Linear Algebra Appl. {\bf 10}, 285 (1975).

\bibitem{rudolph13} F. Verstraete, J. Dehaene, and B. DeMoor, \pra {\bf 64}, 010101 (2001); S. Jevtic, M. Pusey, D. Jennings, and T. Rudolph, arXiv:1303.4724 [quant-ph]. 


\bibitem{zaraket04} S. Hamieh, R. Kobes, and H. Zaraket, \pra {70} 052325 (2004). 

\bibitem{ariano05} G. M. D'Ariano, P. L. Presti, and P. Perinotti, J. Phys. A: Math. Gen. {\bf 38}, 5979 (2005). 

\bibitem{koashi04} M. Koashi and A. Winter, \pra {\bf 69}, 022309 (2004).


\bibitem{du-rank2} M. Shi, W. Yang, F. Jiang, and J. Du, J. Phys. A: Math. Theor. {\bf 44},  415304  (2011). 

\bibitem{yan11} L. Cen, X. Li, J. Shao, and Y. Yan, \pra {\bf 83}, 054101 (2011).



\bibitem{alierratum} M. Ali, A. R. P. Rau, and G. Alber, Phys. Rev. A {\bf 82}, 069902(E) (2010.) 


\bibitem{zero-disc} 
B. Daki\'{c}, V. Vedral, and \v{C}. Brukner, Phys. Rev. Lett. {\bf 105}, 190502 (2010); 
A. Dutta, arXiv:1003.5256 (quant-ph);  
A. Brodutch and D. R. Terno, \pra {\bf 81}, 062103 (2010); 
T. Zhou, J. Cui, and G. L. Long, Phys. Rev. A {\bf 84}, 062105 (2011);
J.-H. Huang, L. Wang, and S.-Y. Zhu, New J. Phys. {\bf 13}, 063045 (2011); 
L. Chen, E. Chitambar, K. Modi, and G. Vacanti, \pra {\bf 83}, 020101(R) (2011).


\bibitem{chrus10} B. Bylicka and D. Chruscinski, \pra {\bf 81}, 062102 (2010).

\bibitem{chrus12} B. Bylicka and D. Chruscinski, Open Syst. Inf. Dyn. {\bf 19}, 1250006 (2012).

\end{thebibliography}
\end{document}